\documentclass[letterpaper,superscriptaddress,pra,floatfix,twocolumn,aps,10pt,nofootinbib]{revtex4-1}

\usepackage{
	amssymb, 
	amsthm, 
	graphicx, 
	xspace,
	overpic,
	ifthen,
	verbatim,
	subfigure, 
	appendix,
	ifthen,
	nicefrac, 
	color, 
	tensor, 
	etoolbox, 
	centernot, 
	bm,
	MnSymbol 
}

\global\long\def\Es{{^{\sharp}}\hspace{-0.5mm}\mathcal{E}}

\newcommand{\branch}{<}
\newcommand{\rec}{\alpha}

\newcommand{\bra}[1]{\langle #1 \rvert  }
\newcommand{\ket}[1]{\lvert #1 \rangle  }

\newcommand{\braket}[2]{  \langle #1 \vert #2 \rangle  }

\newcommand{\matrixelement}[3]{  \langle #1 \vert #2 \vert #3\rangle  }

\newcommand{\ketbra}[2]{  \lvert #1 \rangle \langle #2 \rvert }

\newcommand{\projector}[1]{  \lvert #1 \rangle \langle #1 \rvert }

\newcommand{\abs}[1]{| #1 |} 
\newcommand{\Abs}[1]{\left| #1 \right|}
\newcommand{\aabs}[1]{|| #1 ||} 
\newcommand{\AAbs}[1]{\left| \left| #1 \right| \right|}

\newcommand{\MI}{\ensuremath{\mathcal{I}}}
\newcommand{\Sys}{\ensuremath{\mathcal{S}}}
\newcommand{\Env}{\ensuremath{\mathcal{E}}}
\newcommand{\Frag}{\ensuremath{\mathcal{F}}}
\newcommand{\FragBar}{{\ensuremath{{\overline{\mathcal{F}}}}}}

\newcommand{\Aux}{\ensuremath{\mathcal{X}}}
\newcommand{\Glob}{\ensuremath{\mathcal{H}}}

\newcommand{\mcA}{\ensuremath{\mathcal{A}}}
\newcommand{\mcB}{\ensuremath{\mathcal{B}}}

\DeclareMathOperator{\Tr}{Tr}

\newcommand{\draftmode}{1}    
\newcommand{\notetoself}[1]{\ifnum \draftmode=1 {\color[rgb]{0,0,0.8} [#1]} \fi}  
\newcommand{\cuttext}[1]{\ifnum \draftmode=1 {\color[rgb]{0,0.5,0} [#1]} \fi}  
\newcommand{\warntext}[1]{\ifnum \draftmode=1 {\color[rgb]{0.9,0.6,0} #1} \else {#1} \color{black} \fi}

\pdfoutput=1

\renewcommand{\draftmode}{0} 

\newcommand{\widewidthfactor}{0.96} 

\newcommand{\DD}{\mathfrak{D}}  
\newcommand{\CF}{\mathfrak{C}}  

\newcommand{\CN}{\mathrm{\textsc{CNOT}}}
\newcommand{\CNOT}{\textsc{CNOT}}

\begin{document}



\title{The Objective past of a quantum universe: Redundant records of consistent histories}
\date{\today}
\author{C.~Jess~Riedel}
\email{jessriedel@gmail.com}
\affiliation{Perimeter Institute for Theoretical Physics, Waterloo, Ontario N2L 2Y5, Canada}
\affiliation{IBM Watson Research Center, Yorktown Heights, NY 10598}
\affiliation{Theoretical Division, LANL, Los Alamos, New Mexico 87545}
\author{Wojciech~H.~Zurek}
\affiliation{Theoretical Division, LANL, Los Alamos, New Mexico 87545}
\author{Michael~Zwolak}
\affiliation{Department of Physics, Oregon State University, Corvallis, OR 97331}


\begin{abstract}
Motivated by the advances of quantum Darwinism and recognizing the role played by redundancy in identifying the small subset of quantum states with resilience characteristic of objective classical reality, we explore the implications of redundant records for consistent histories. The consistent histories formalism is a tool for describing sequences of events taking place in an evolving closed quantum system. A set of histories is consistent when one can reason about them using Boolean logic, i.e., when probabilities of sequences of events that define histories are additive. However, the vast majority of the sets of histories that are merely consistent are flagrantly non-classical in other respects. This {\it embarras de richesses} (known as the set selection problem) suggests that one must go beyond consistency to identify how the classical past arises in our quantum Universe. The key intuition we follow is that the records of events that define the familiar objective past are inscribed in many distinct systems, e.g., subsystems of the environment, and are accessible locally in space and time to observers. We identify histories that are not just consistent but redundantly consistent using the partial-trace condition introduced by Finkelstein as a bridge between histories and decoherence. The existence of redundant records is a sufficient condition for redundant consistency. It selects, from the multitude of the alternative sets of consistent histories, a small subset endowed with redundant records characteristic of the objective classical past. The information about an objective history of the past is then simultaneously within reach of many, who can independently reconstruct it and arrive at compatible conclusions in the present.
\end{abstract}


\maketitle

\section{Introduction}
\label{sec:intro}

``Into what mixture does the wavepacket collapse?''  This is the \emph{preferred basis problem} in quantum mechanics \cite{Zurek1981}. It launched the study of decoherence \cite{Zurek1982,Joos1985}, a process central to the modern view of the quantum-classical transition \cite{Zeh1973, zurek1991decoherence,zurek2003a,JoosText,SchlosshauerText, *schlosshauer2014quantum--classical,ZurekPT2014}. The preferred basis problem has been solved exactly for so-called \emph{pure decoherence} \cite{Zurek1981,zwolak2014amplification}. In this case, a well-defined \emph{pointer basis} \cite{Zurek1981} emerges whose origins can be traced back to the interaction Hamiltonian between the quantum system $\Sys$ and its environment $\Env$ \cite{Zeh1973, Zurek1981, Zurek1982}.  An approximate pointer basis exists for many other situations (see, e.g., Refs.~\cite{anglin1996decoherence,*dalvit2005predictability,Kubler1973, Zurek1993a, Gell-Mann1993, Gell-Mann2007, Halliwell1998, Paz1999}).

The consistent (or decoherent) histories framework \cite{Griffiths1984, OmnesText, GriffithsText, HalliwellReview} was originally introduced by Griffiths. It has evolved into a mathematical formalism for applying quantum mechanics to completely closed systems, up to and including the whole universe.  It has been argued that quantum mechanics within this framework would be a fully satisfactory physical theory only if it were supplemented with an unambiguous mechanism for identifying a preferred set of histories corresponding, at the least, to the perceptions of observers \cite{Dowker1995, Dowker1996, Kent1996,Kent1997a,Kent1997b, KentRemarks, okon2014measurements, okon2015consistent} (but see counterarguments \cite{griffiths1998comment,griffiths1998choice,griffiths2013consistent,griffiths2013epistemic,griffiths2014new,griffiths2015consistent}). This would address the Everettian \cite{everett1957relative} question: ``What are the branches in the wavefunction of the Universe?'' This defines the \emph{set selection problem}, the global analog to the preferred basis problem.

It is natural to demand that such a set of histories satisfy the mathematical requirement of \emph{consistency}, i.e., that their probabilities are additive. The set selection problem still looms large, however, as almost all consistent sets bear no resemblance to the classical reality we perceive \cite{Paz1993,Zurek1993b, Gell-Mann1998}.   Classical reasoning can only be done \emph{relative} to a single consistent set \cite{griffiths1998choice, griffiths2013consistent, GriffithsText}; simultaneous reasoning from different sets leads to contradictions \cite{Gell-Mann1994, Omnes1992, Dowker1995, Dowker1996, Kent1996}. A preferred set would allow one to unambiguously compute probabilities\footnote{We take Born's rule for granted, putting aside the  question of whether it should be derived from other principles \cite{gleason1957measures, everett1957relative, hartle1968quantum, zurek2003environment, zurek2011entanglement, Zurek2005, ZurekPT2014, Sebens2014, nenashev2014quantum, *nenashev2016why} or simply assumed.  That issue is independent of (and cleanly separated from) the topic of this paper.} for all observations from first principles, that is, from (1) a wavefunction of the Universe and (2) a Hamiltonian describing the interactions.

To agree with our expectations, a preferred set would describe macroscopic systems via coarse-grained variables that approximately obey classical equations of motion, thereby constituting a ``quasiclassical domain'' \cite{Gell-Mann1990b, Gell-Mann1993, Gell-Mann1994, Dowker1996, Kent1996, tegmark2015consciousness}.  Various  principles for its identification have been explored, both within the consistent histories formalism \cite{Gell-Mann1990b, Griffiths1996, Anastopoulos1997, Kent1997b, Kent1998, Gell-Mann1998, Brun1999, Kent2000, Brun2003, Gell-Mann2007} and outside it \cite{debroglie1928nouvelle, bohm1952suggested, bousso2012multiverse, Kent2012, kent2014solution}.   None have gathered broad support.

It has long been recognized that records may play a key role in identifying the quasiclassical domain \cite{ZurekTime, Zurek1982, Gell-Mann1990b, Gell-Mann1993, Halliwell1999, Dodd2003}. We say a record of one system exists in another system when a measurement of the latter can provide information about the state of the former, a property that can be quantified by the mutual information between the two \cite{Zurek2009}. When a system decoheres, records about its state (in the pointer basis) are created in the environment.  Intuitively, \emph{sequences} of records created at different times -- either deliberately by an apparatus or accidentally, but inevitably, by a natural environment -- will similarly decohere a preferred set of histories, analogous to the preferred set of pointer states.  Indeed, for any consistent set of histories of a pure global state, Gell-Mann and Hartle pointed out that, at least formally, records of the history must exist in the state as a whole \cite{Gell-Mann1993}.  Here, ``global'' refers to everything: both the system and its environment, which can encompass the whole universe.  This argument guarantees only the existence of \emph{one} record, however. Without a stronger principle for identifying a preferred set (which Gell-Mann and Hartle explore), this formal record will generally correspond to a highly complicated observable involving the entire global state, completely inaccessible to a realistic observer inside the Universe described by that global state. (An example illustrating the delocalized nature of such records can be found in the Appendix.) Moreover, when the global state is mixed, there need be no record at all.

Establishing realistic records will require something more. It has been noted that the decoherence of a quantum system is connected to the production of a record about that system in the environment \cite{Zurek1981, Zurek1982, Halliwell1999}, but the ability of \emph{many} observers to infer the state of the system by sampling an environment is predicated on \emph{redundant records} -- a property of  some global states that does not follow merely from decoherence (or consistency).  

\emph{Quantum Darwinism} \cite{Zurek2000, Zurek2009, horodecki2015quantum, brandao2015generic} is a paradigm for describing and quantifying what distinguishes such states awash in the enormous sea of Hilbert space.  It has been successfully applied to several prototypical models of decoherence \cite{Blume-Kohout2005, Blume-Kohout2008, Paz2009, Zwolak2009, Zwolak2010, Riedel2010, Riedel2011, Riedel2012, korbicz2014objectivity-a, zwolak2014amplification}, but models with non-trivial histories -- those featuring multiple events at different times, each with multiple outcomes -- have not yet been investigated.

Some mathematical way of isolating records within the universe is necessary for distinguishing different copies of a record from each another.  Indeed, real-world observers are constrained to acquire information via measurements that are local in space and time. This is why the photon environment, rather than the air environment (which is typically more effective than photons in decohering macroscopic objects), is responsible for most of the information we acquire: Air molecules interact with each other, delocalizing the data they have obtained \cite{Riedel2012}, and preventing the information they collectively hold from being accessible via local measurements. In contrast, photons scatter from objects of interest but not from each other. This allows multiple observers to find out about the world via local measurements on small fractions of the photon environment.  We simply assume in what follows that \emph{some} tensor structure dividing the universe into parts exists.  One might imagine these to be atoms, phonon modes, or cubic spatial volumes.

In this work, we extend the concept of redundant records to histories using the notion of \emph{partial-trace consistency}. Two decades ago, Finkelstein introduced the mathematical condition of partial-trace consistency (or \emph{partial-trace decoherence}) to identify when the consistency of a set of histories of a system is due to that system's interaction with an environment \cite{Finkelstein1993} (see also Refs.~\cite{ZurekTime, Zurek1993b}).   For a pure global state, partial-trace consistency precisely defines the idea that records exist -- not just somewhere -- but \emph{in the environment}.  This provides a key link among consistency, decoherence, and localized records.

As we will show, partial-trace consistency (which we generalize) allows us to specify which \emph{parts} of the environment are responsible for consistency.  For realistic systems that are decohered by large environments, we expect the consistency to be highly redundant in the sense that the environment will have many small parts that are each individually sufficient to decohere the system.  Likewise, we expect records of the history to be widely available locally in space and time from many small parts of the environment. 

Redundancy of the records requires resilience of the original states that define events (sequences of which constitute histories). These states must survive ``copying'' by the environment. To be repeatedly copied, events must be distinguishable, i.e., they must correspond to orthogonal states~\cite{Zurek2007} or, when the system is macroscopic, to orthogonal subspaces~\cite{zurek2013wave-packet} of the system's Hilbert space.

With all this in hand, we can rigorously define the idea that a history is recorded redundantly, thereby establishing the objectivity of a set of histories in the sense that many independent observers can all agree on the history by accessing only small, disjoint subsets of the environment. Our everyday macroscopic observations are objective in this sense. Thus, uniqueness results from quantum Darwinism \cite{Ollivier2004,Ollivier2005,horodecki2015quantum, brandao2015generic, zwolak2013complementarity} can be used to narrow down the sets of consistent histories that describe the quasiclassical domain.  This opens a new line of attack on the set selection problem\footnote{Note that Griffiths argues that the various incompatible consistent sets of histories -- although possibly of differing utility to a particular observer -- are fundamentally equally valid descriptions of the universe.  This is compared to the different thermodynamic coarse grainings that are equally valid descriptions of a classical system, although the classical situation differs importantly from the quantum case in that a unique fine graining compatible with all coarse grainings exists \cite{griffiths2013epistemic, griffiths2014new}.  We are interested in histories relevant for observers who access the past through its records here and now. Thus, in the framework suggested by Griffiths, the restriction of spatiotemporal locality greatly alleviates the set selection problem.} and shows how our experience of persistent classicality arises within a fundamentally quantum universe.  


The outline of this work is as follows. Sections \ref{sec:decoherence} and \ref{sec:consistent-histories} review decoherence and the consistent histories framework, respectively.  Readers familiar with these ideas can skip those sections after noting our choice of notation. We describe partial-trace consistency and redundant records for histories in Section \ref{sec:rptconsis}, expanding on the results of Ref.\ \cite{Finkelstein1993}.  In particular, we prove an identity linking the fidelity of two conditional states of a subsystem and the corresponding partial-trace decoherence functional. In Section \ref{sec:CNOT-example} we apply these ideas to a simple model of a two-state system monitored by an environment through ``controlled-not'' interactions.  Finally, we conclude in Section \ref{sec:thisdiscussion}.  The connection to quantum Darwinism will be completed in a forthcoming companion paper \cite{riedel2013objective2}.


We take $\hbar = 1$. Although not written explicitly, states like $\ket{\psi_\alpha}$ and $\rho^\Frag$ have implicit dependence on $t$.   If we wish to specify that they are to be evaluated at a particular time $t_*$, then we use the notation $\rho \vert_{t_*} = U_{t,t_*}\rho U_{t,t_*}^\dagger$, where $U_{t',t''}$ is the unitary that evolves from $t'$ to $t''$.  Initial states will be tagged with a null superscript (e.g., $\rho^0$, $\ket{\psi^{\Sys,0}}$) and will be considered fixed, as will certain basis states  (e.g., $\ket{0}, \ket{1}$).  History projectors, $P(t_m) = U_{t,t_m}^\dagger P U_{t,t_m}$, have explicit dependence on $t_m$ and implicit dependence on $t$.  (As seen below, these two times are associated with the relevant event in the past and with its consequences in the present, respectively.)  Similarly, decoherence factors ($\Gamma_{s s^\prime}$), class operators ($C_\alpha$), decoherence functionals [$\DD_{\Frag}(\alpha,\beta)$], and information theoretic quantities (e.g., $\MI_{\Sys:\Frag}$, $H_\Env$) have implicit dependence on $t$.  Symbols like $\Sys$ will be used to both label a quantum system and also to denote its associated Hilbert space.  We put a hat on the Hamiltonian $\hat{H}$ to distinguish it from von Neumann entropies (e.g., $\hat{H}_\Sys$ versus $H_{\Sys}$).

\section{Decoherence and Quantum Darwinism}
\label{sec:decoherence}




Observers and other macroscopic quantum system do not exist in isolation, but are virtually always immersed in a large environment. The resulting decoherence has profound implications for their behavior \cite{zurek1991decoherence,zurek2003a,JoosText, SchlosshauerText,*schlosshauer2014quantum--classical,ZurekPT2014}.  Below we briefly review some basic decoherence concepts.

\subsection{Pure decoherence}
\label{subsec:puredecoh}

Suppose a system $\Sys$ starts in an arbitrary pure state
\begin{align}
\ket{\psi^{\Sys,0}} = \sum_s c_s \ket{s}
\end{align}
in a fixed basis $\ket{s}$ and it is coupled to an environment $\Env$ initially in some state $\rho^{\Env_,0}$. The system $\Sys$ is said to undergo \emph{pure decoherence} \cite{zwolak2014amplification} by an environment when the evolution takes the form of a controlled unitary
\begin{align}
\label{eq:puredecoh}
e^{-i t \hat{H}} = \sum_s \ket{s} \bra{s} \otimes U_s,
\end{align}
where the $U_s$ are arbitrary unitaries governing the evolution of $\Env$ conditional on the state $\ket{s}$ of $\Sys$.  In this case, the set $\{\ket{s}\}$ forms an unambiguous \emph{pointer basis} for the system $\Sys$ \cite{Zurek1981}. The global state is then
\begin{align}
\rho = \sum_{s,s'} c_s c_{s'}^* \ket{s} \bra{s'} \otimes U_s \rho^{\Env,0} U_{s'}^{\dag}
\end{align}
and the density matrix of $\Sys$ is
\begin{align}
\rho^\Sys = \sum_{s,s'}  \Gamma_{ss'} c_s c_{s'}^* \ket{s} \bra{s'}
\end{align}
where the \emph{decoherence factors} are
\begin{align}
\label{eq:puredecohfactor}
\Gamma_{s s'} = \Tr [U_s \rho^{\Env,0} U_{s'}^{\dag}] = \Gamma_{s' s}^*.
\end{align}
When $\Gamma_{s s'}$ vanishes for all $s \neq s'$, we say that the system has been fully decohered by the environment; it takes the diagonal form in the pointer basis
\begin{align}
\rho^\Sys = \sum_{s}  \Abs{c_s}^2 \ket{s} \bra{s}.
\end{align}


\subsection{Records, mixtures, and purifications}
\label{subsec:records}

A \emph{record} of the state of $\Sys$ exists in another system -- e.g.,  an environment $\Env$ -- when a measurement on that system can provide information about the state of $\Sys$ (whether or not a measurement is actually made).
For instance, if the global state of 
$\Sys \Env$ is
\begin{align}
\ket{\psi} = \sum_{s}  c_s \ket{s} \ket{e_s} ,
\end{align}
with $\{\ket{e_s}\}$ an orthonormal set in $\Env$, then we say $\Env$ has a record of the state $\ket{s}$ of $\Sys$.

Given that the von Neumann measurement scheme \cite{neumann1932mathematische} is the prototypical example of pure decoherence, one might expect that pure decoherence (of the pointer basis $\{\ket{s}\}$ by $\Env$) necessarily leads to records (about $\{\ket{s}\}$ in $\Env$).  In fact, for a general mixed initial state $\rho^{\Env,0}$, the decoherence of $\{\ket{s}\}$ by $\Env$ does not guarantee any such information.  The crucial difference between mere decoherence and the existence of records is encoded in the distinguishability of the conditional states of the environment, $\rho_s^\Env = U_s \rho^{\Env,0} U_s^{\dag}$.  For an observer to be able to clearly discriminate between the various pointer states by making a measurement on $\Env$, the conditional states $\rho^\Env_s$ must be mutually orthogonal.  This automatically implies that decoherence factors $\Gamma_{s s^\prime}$ vanish for $s \neq s^\prime$ (records imply decoherence) but the converse is not true, as we will now see.

Take $\Sys$ to be a qubit initially in the superposition $\ket{\psi}_\Sys = \ket{0}_\Sys + \ket{1}_\Sys$ and let $\Env$ be a fully mixed qubit $\rho^{\Env,0} = \mathrm{I}$.  (We ignore normalization in the rest of this subsection.) Without loss of generality, we can postulate a non-interacting auxiliary system $\Aux$ which purifies $\Env$ using the joint initial state
\begin{gather}
\ket{EX}_{\Env \Aux} = \ket{0}_\Env \ket{0}_\Aux+\ket{1}_\Env \ket{1}_\Aux, \\
\rho^{\Env,0} = \ket{0}_\Env \bra{0} +\ket{1}_\Env \bra{1}.
\end{gather}
We can produce pure decoherence using a controlled-not (\CNOT) gate
\begin{align}\begin{split}\label{eq:CSHIFT}
U^{\Sys \Env}_{\CN} = \ket{0}_\Sys \bra{0} \otimes I^\Env + \ket{1}_\Sys \bra{1} \otimes \big(\ket{0}_\Env \bra{1} +\ket{1}_\Env \bra{0} \big)
\end{split}\end{align}
which flips the state of $\Env$ when $\Sys$ is in the state $\ket{1}_\Sys$.  This yields
\begin{align}\begin{split}\label{eq:decoh-no-record}
U^{\Sys \Env}_{\CN} \ket{\psi}_\Sys \ket{EX}_{\Env \Aux} &= \ket{0}_\Sys \Big[\ket{0}_\Env \ket{0}_\Aux+\ket{1}_\Env \ket{1}_\Aux \Big]\\
 & \, \quad +  \ket{1}_\Sys \Big[\ket{1}_\Env \ket{0}_\Aux + \ket{0}_\Env \ket{1}_\Aux \Big].
\end{split}\end{align}
There is no record in $\Env$ about the pointer basis $\{ \ket{0}_\Sys, \ket{1}_\Sys \}$ even though this is the basis in which $\Sys$ is decohered.  The distinction between $\ket{0}_\Sys$ and $\ket{1}_\Sys$ has been recorded in the expanded environment $\Env \otimes \Aux$, but this information is not accessible in just $\Env$.  Rather, it is encoded in the \emph{correlation} between $\Env$ and $\Aux$: When $\Sys$ is up, $\Env$ and $\Aux$ point in the same direction.  When $\Sys$ is down, they point in opposite directions.  Looking at \emph{just} $\Env$ (or just $\Aux$), however, tells one nothing about the pointer basis.  More generally, when the initial state of $\Env$ is partially mixed, the ability of $\Env$ to record the pointer states of $\Sys$ can be analyzed using methods from communication theory by treating $\Env$ as a noisy channel \cite{Zwolak2009, Zwolak2010,zwolak2014amplification,QCBspins}.

We can get a final state with the same form as Eq.~\eqref{eq:decoh-no-record} using a pure global state by allowing inter-environmental interactions. For instance, starting in the product initial state $\ket{\psi}_\Sys \ket{0}_\Env \ket{0}_\Aux$ and applying a \CNOT{} gate to both $\Env$ and $\Aux$ (conditional on $\Sys$) yields the GHZ state:
\begin{align}\begin{split}\label{eq:decoh-ghz}
U^{\Sys \Aux}_{\CN} U^{\Sys \Env}_{\CN} \ket{\psi}_\Sys \ket{0}_\Env \ket{0}_\Aux = \ket{0}_\Sys \ket{0}_\Env \ket{0}_\Aux + \ket{1}_\Sys \ket{1}_\Env  \ket{1}_\Aux.
\end{split}\end{align}
This is pure decoherence with a product global state, which is essentially guaranteed to yield redundant records \cite{brandao2015generic,zwolak2014amplification,QCBspins}.  At this point, $\Env$ and $\Aux$ both have a complete record of $\Sys$.  A local interaction between $\Env$ and $\Aux$ taking $\ket{00} \to \ket{00}+\ket{11}$ and $\ket{11} \to \ket{10}+\ket{01}$ can recover the state in the form of Eq.~\eqref{eq:decoh-no-record}.

The localization of macroscopic bodies by blackbody illumination provides a useful example of decoherence by a mixed environment.  Photons from concentrated sources like the sun can produce records despite being partially mixed, but isotropic thermal illumination, such as in a uniform oven, fully decoheres the system without yielding localized records \cite{Riedel2011}.



\newcommand{\introdfactor}{0.73} 

\begin{figure} [bt]
	\centering
	\includegraphics[width=8cm]{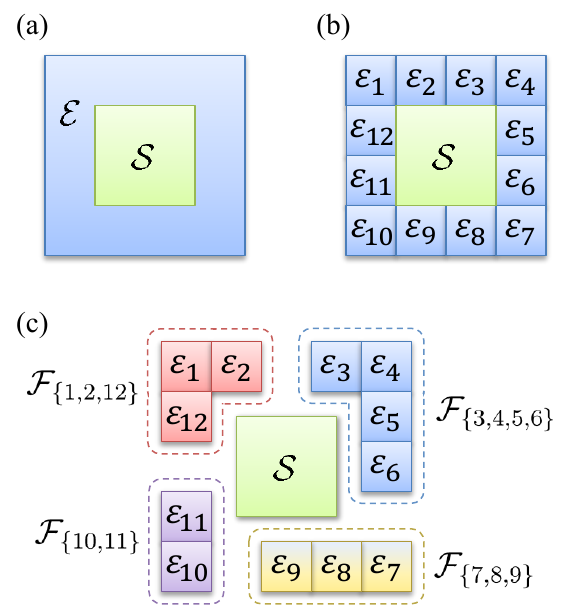}
	\caption{(a) In the original decoherence paradigm, the system $\Sys$ is immersed in a monolithic environment $\Env$ which is traced over. (b)  Quantum Darwinism is based on recognizing the natural decomposition of the environment into parts $\varepsilon_k$ ($k=1,\ldots,\Es$), where $\Es$ is the total number of parts of the environment. (c)  Fragments $\Frag_{\{ n \}}$ partition $\Env$ into multiple pieces that may be accessed by different observers.}
	\label{fig:introd}
\end{figure}

\subsection{Records in the fragments of the environment}
\label{subsec:fragmentation}       %

The location and accessibility of records will be of paramount importance.  For this reason, we will be interested in fragments of the environment formed by proper subsets of its subsystems, as depicted in Figure \ref{fig:introd}.  In this case, evolution leading to pure decoherence is induced by a unitary in the form of
\begin{align}
e^{-i t \hat{H}} = \sum_s \projector{s} \otimes \bigotimes_{k=1}^{\Es} U_s^{k},
\end{align}
where the environment is composed of $\Es$ parts that are initially uncorrelated,
\begin{align}
\rho^0 = \rho^{\Sys,0} \otimes  \left[ \bigotimes_{k=1}^{\Es} \rho^{k,0} \right],
\end{align}
and do not interact with one another [thus avoiding complications illustrated by \eqref{eq:decoh-no-record} and the ensuing discussion].

A \emph{fragment} $\Frag$ is a subset of the environment components. As context will make clear, we use $\Frag$ to designate either the subset of the environment component labels, i.e., $\Frag \subset \{1, \ldots, \Es\}$, or as the Hilbert space of that subset of environment components. We also define the fragment's complement  $\FragBar$ such that $\Env = \Frag \otimes \FragBar$. This allows us to isolate a fragment's contribution to the decoherence factor: $\Gamma_{s s'} = \Gamma_{s s'}^\Frag \Gamma_{s s'}^{\FragBar}$ where
\begin{align}
\label{eq:partialdecohfactor}
\Gamma_{s s'}^\Frag = \Tr [U_s^{\Frag} \rho^{\Frag,0} U_{s'}^{\Frag \dag}] = \Gamma_{s' s}^{\Frag *}.
\end{align}

Equation \eqref{eq:decoh-ghz} features two records of $\Sys$, one in $\Env$ and one in $\Aux$.  As discussed earlier, however, we expect real-life quantum systems to have many records, taking the form of \emph{branching states} \cite{blume-kohout2006quantum, Riedel2012}:
\begin{align}
\label{eq:ghz}
\ket{\psi} \simeq \sum_s c_s  \ket{s} \ket{F_s} \ket{F^{\prime}_s} \ket{F^{\prime\prime}_s} \cdots.
\end{align}
These are GHZ-like states, with each term in the sum corresponding to a ``branch'' in the global wavefunction.  We do not necessarily expect this idealized branching structure on the microscopic level.   (Individual particles may be entangled within a branch, e.g., in spin singlet states.)  Our primary concern, though, is with tracking the history of macroscopic variables, for which large fragments $\Frag_i$ will often approximately take the form above and the redundant records will exist.

More generally, when the environment subsystems interact, as do air molecules, they will entangle, scrambling the information about the pointer states, so it will be no longer contained in natural (local) environment fragments \cite{Riedel2012}. For the photon environment we find most useful for acquiring information this is usually not the case: Photons scatter off the systems of interest, but do not interact with one another. Intermediate environments that scramble part of the information are also possible \cite{Riedel2012}.

\subsection{Quantum Darwinism}
\label{subsec:Witness}       %

Quantum Darwinism recognizes that observers acquire information about the states of systems of interest in our Universe indirectly, by monitoring fragments of the environment that decoheres these systems \cite{Zurek2000, Zurek2009, horodecki2015quantum, brandao2015generic}. For humans, the photon environment is usually responsible for most of the acquired information. 

Mutual information between the system and a fragment of the environment is given by:
$$I(\Sys:\Frag)=H_\Sys+H_\Frag-H_{\Sys,\Frag}$$
where $H_\Sys,~H_\Frag$, and $H_{\Sys,\Frag}$ are the von Neumann entropies of $\Sys$, $\Frag$, and $\Sys\Frag$ taken jointly. $I(\Sys:\Frag)$ is a measure of how much information is in the record -- how much $\Sys$ and $\Frag$ know about each other. A fragment $\Frag$ contains an approximate record of $\Sys$ when 
$$ I(\Sys:\Frag)=(1-\delta)H_\Sys$$
that is, when one can reduce the entropy of the system to just $\delta H_\Sys$ by a measurement\footnote{We note that a ``measurement'' or an ``observation'' used to access the record in $\Frag$ can be discussed (and are meant here) in an operational, interpretation-independent fashion. All that is needed is a suitable interaction between the fragment $\Frag$ and the observer or an apparatus that we can denote by $\cal A$. When this interaction results in mutual information between $\Sys$ and $\cal A$, then (part of) the information that $\Frag$ had about $\Sys$ has been passed onto $\cal A$. Interesting questions that can be posed about this imperfect copying process (see, e.g., Refs.~\cite{Zurek2007,zurek2013wave-packet}), but they are beyond the scope of this paper.} on $\Frag$. The {\it information deficit} $\delta$ quantifies the quality of the record.  Records of $\Sys$ are redundant when there are many disjoint fragments of the environment that can provide this information. The number of such fragments defines the \emph{redundancy}, a quantity which can be huge and generally depends only logarithmically on the information deficit $\delta$ \cite{zwolak2014amplification, Blume-Kohout2005, Blume-Kohout2008, Paz2009, Zwolak2009, Zwolak2010, Riedel2010, Riedel2011, Riedel2012, Ollivier2004, Ollivier2005, QCBspins}.

When many records of the instantaneous state of an evolving system are deposited in the environment at different times, observers can use them to reconstruct the system's history. Intuitively, such histories are objective, as the redundant records make the history accessible to many, who -- when they compare their accounts -- will reach consistent conclusions. It also turns out to guarantee that the histories are redundantly consistent, in a way that will be made precise.

The quantum Darwinism view of histories is a natural extension of the same approach that was successfully employed to account for objective existence of states in our quantum Universe. However, dealing with histories is more technically challenging than with just states. This paper begins to address some of the most obvious aspects of this challenge: The relation between the consistency of histories and creation of the multiple records of events that constitute them. As we have already pointed out, and as was also the case for states, the decohering environment does not always make such records accessible. Redundant decoherence, even in the absence of the accessibility of records, assures the additivity of the probabilities of histories -- i.e., consistency. As we will see below, though, redundancy constrains the histories and thus sheds new light on the set selection problem.  A forthcoming companion paper will review the quantum Darwinism framework in detail and, using concepts introduced herein, will define a notion of redundancy of records of a history that reduces to the quantum Darwinian view of states~\cite{riedel2013objective2}.

\section{Consistent histories}
\label{sec:consistent-histories}       %

In this section, the consistent histories framework \cite {Griffiths1984} is briefly summarized. See Refs.\ \cite{HalliwellReview, OmnesText, GriffithsText} and references therein for a broad review.

\subsection{The consistent histories framework}
\label{subsec:ch-framework}       %


At a fixed time $t$, alternative outcomes within a closed quantum system are represented by a complete set of orthogonal projection operators $\{ P_a \}$,
satisfying
\begin{align}
\sum_a P_a &= I,  \\
P_a P_b &= \delta_{ab} P_a, \\
P_a^\dagger &= P_a.
\end{align}
The probabilities of these outcomes for a given state $\rho$ are given by
\begin{align}
p_a = \Tr [P_a \rho P_a^\dag] =  \Tr [P_a \rho].
\end{align}
In the case of a pure global state $\rho = \projector{\psi}$, the probability is just the squared norm of the conditional state $\ket{\psi_a} = P_a \ket{\psi}$.


A \emph{history} is a sequence\footnote{For simplicity, we are restricting our attention to \emph{homogeneous} histories \cite{isham1994quantum} (also known as ``histories with chain form''), i.e., those histories which can be expressed as a single time-ordered string of Heisenberg-picture projections.  We do not address \emph{inhomogeneous} histories formed by summing homogeneous ones together, which may become important in relativistic settings.  Also for simplicity, we do not consider \emph{branch-dependent} histories \cite{Gell-Mann1993} for which projectors at later times are chosen conditional on projectors at earlier times.} of several outcomes at different times $t_m$, $m=1,\ldots,M$.  It defines a \emph{class operator}\footnote{The term ``class'' for these operators reflects the fact that the projectors composing them are generally multi-dimensional and, especially in the context of a path-integral formulation, may pick out a certain class of possible Feynman histories through position space \cite{hartle1995spacetime}.  That is, the class is the coarse-grained set, and the many fine-grained histories are the members of the class.  However, we do not assume here that the projectors forming histories are diagonal in the position basis.} of the form\footnote{Note that our definition lacks the leading unitary found in Finkelstein's convention \cite{Finkelstein1993} for reasons explained in Sec. \ref{sec:rptconsis}.}
\begin{align}
\label{eq:classop}
C_{\alpha} &=  P_{a_M}^{(M)} (t_M) \cdots P_{a_1}^{(1)}(t_1),
\end{align}
where $\alpha = (a_1, \ldots, a_M)$.  Above,
\begin{align}
P_{a_m}^{(m)} (t_m) = U_{t,t_m}^\dag P_{a_m}^{(m)} U_{t,t_m}
\end{align}
are the time-evolved projectors and $U_{t',t''} = e^{-i(t''-t')H} = U_{t'',t'}^\dag$ is the unitary which evolves from $t'$ to $t''$.  The set $\{ P_{a_m}^{(m)} \}$ is a complete set of projectors, indexed by $a_m$, for each fixed value of $m$.  A complete set of histories ranges over all values of $\alpha$, and the corresponding class operators obey
\begin{align}
\label{eq:classopres}
\sum_{\alpha} C_{\alpha} = I.
\end{align}

Importantly, both the projectors $P_{a_m}^{(m)} (t_m)$ and the class operators $C_{\alpha}$ have (like $\rho$) implicit dependence on an unsubscripted time $t$.   As will become clear in Section \ref{sec:rptconsis}, the variable $t$ will range over different times both before and after physical records are created.  On the other hand, the $t_m$ are the fixed times in the past when historical events occurred.

One would like to assign probabilities to histories according to
\begin{align}
\label{eq:hisprob}
p_\alpha = \Tr \left[ C_{\alpha} \rho \, C_{\alpha}^{\dag} \right]  .
\end{align}
In the case of a pure state $\ket{\psi}$, this would again be just the squared norm of the conditional state $\ket{\psi_\alpha} = C_\alpha \ket{\psi}$.  Although such numbers necessarily are real, positive, and sum to unity, they do not obey the probability sum rules.  That is, if we use the definition \eqref{eq:hisprob} then it is not generally true that the probability assigned to the sum of two mutually exclusive histories equals the sum of the probabilities for the two individually\footnote{Our use of the logical operators $\wedge$ (``and'') and $\vee$ (``or'') emphasizes that the consistent histories formalism is the natural embedding of classical logic in quantum mechanics \cite{Griffiths1984,Omnes1992,omnes1988logical,*omnes1988logical-a,*omnes1988logical-b,*omnes1989logical,isham1995quantum,GriffithsText}.}:
\begin{align}
\label{eq:probsum}
p_{\alpha \vee \beta} = p_\alpha + p_\beta,
\end{align}
where $C_{\alpha \vee \beta} = C_\alpha + C_\beta$ is a \emph{coarse graining} of $C_\alpha$ and $C_\beta$.  ($C_\alpha$ and $C_\beta$ are \emph{fine grainings} of $C_{\alpha \vee \beta}$, and the operations of coarse and fine graining of histories form a partial order \cite{Gell-Mann1993}.) As pointed out by Griffiths \cite{Griffiths1984}, who used it to motivate the consistent histories approach, the failure of this probability sum rule to hold is a manifestation of quantum interference, which prevents the indiscriminate application of classical logic to quantum systems.

\begin{figure} [bt]
  \centering
  \includegraphics[width=\widewidthfactor\columnwidth]{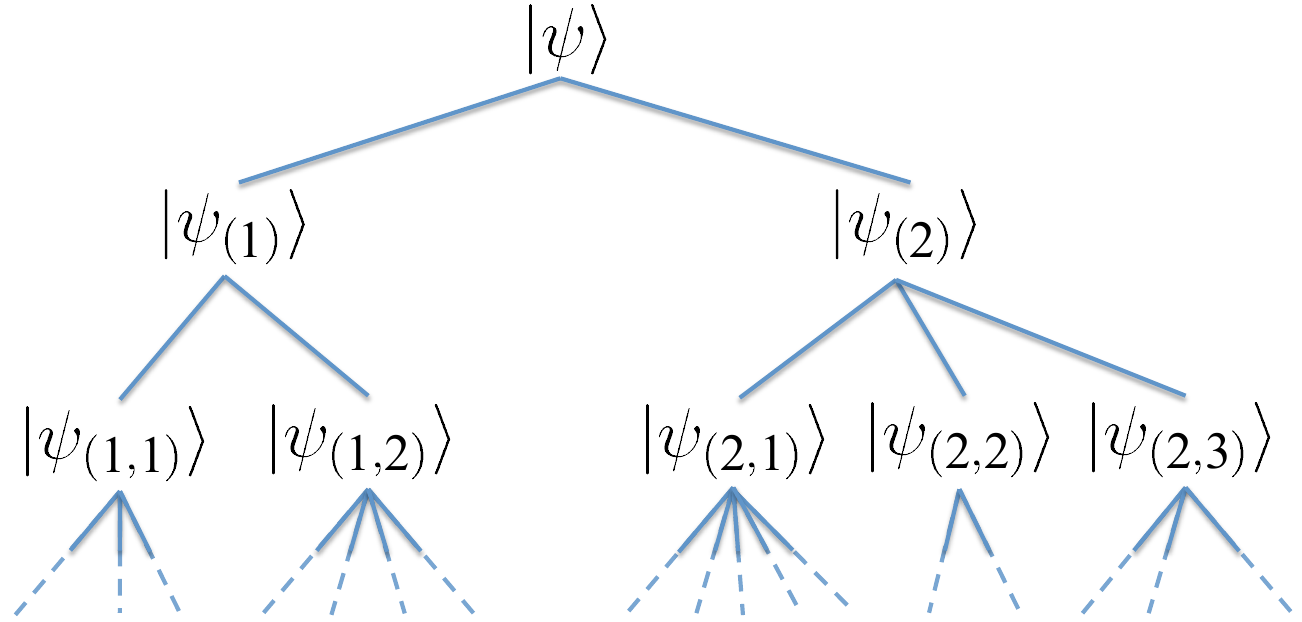}
  \caption{When the global state of a consistent set of histories $\{ \alpha \}$ is pure, $\rho = \projector{\psi}$, the branches $\ket{\psi_\alpha} = C_\alpha \ket{\psi}$ take on a simple tree structure.  The top (zeroth) level is just the global state $\ket{\psi}$ while the first level contains those branches formed by the projectors $P^{(1)}_{a_1}$ at time $t_1$: $\ket{\psi_{(a_1)}} = C_{(a_1)} \ket{\psi} = P^{(1)}_{a_1} (t_1) \ket{\psi}$.  Across each level, the branches  are mutually orthogonal and together sum to $\ket{\psi}$.  Each branch is also the sum of those branches connected directly below it.  This structure is a result of consistency, $\DD (\alpha, \beta) = 0 = \braket{\psi_\alpha}{\psi_\beta}$ for $\alpha \neq \beta$.  It does not follow just from Eq.~\eqref{eq:classop}.}
  \label{fig:purestatepartialorder}
\end{figure}

We therefore require that a set of histories obey the condition
\begin{align}
\label{eq:meddecoh}
\DD(\alpha, \beta) = 0 , \qquad  \alpha \ne \beta,
\end{align}
where
\begin{align}
\DD(\alpha, \beta) = \Tr \left[ C_{\alpha} \rho \, C_{\beta}^{\dag} \right]
\end{align}
is the \emph{decoherence functional}.  If this condition is satisfied, we say that the histories are \emph{consistent}.\footnote{There are reasons to consider other conditions \cite{Goldstein1995, Hartle2004, Gell-Mann1998, Halliwell2009} besides Eq.~\eqref{eq:meddecoh}.  In particular, the weaker requirement that $\mathrm{Re} \; \DD(\alpha, \beta) = 0$  for  $\alpha \ne \beta$ is often known as `consistency', since it allows for consistent probabilities to be defined.  Such probabilities will not be robust under the composition of subsystems \cite{Diosi1994,Diosi2004}, so this condition is usually rejected as too weak. In this article, we always mean Eq.~\eqref{eq:meddecoh} when we say ``consistency''.}  Equation \eqref{eq:meddecoh} has often been called \emph{medium decoherence} \cite{Gell-Mann1990b}.  We resist this practice and reserve ``decoherence'' for the physical process predicated on a system-environment distinction as discussed in Section \ref{subsec:consisdecoh}.

We can also define for any subsystem $\mcA$ (where $\Glob = \mcA \otimes \mcB$) the normalized state of $\mcA$ conditional on $\alpha$:
\begin{align}\begin{split}
\label{eq:condmatforhis}
\rho_\alpha^\mcA = \frac{\Tr_{\mcB} \left[ C_\alpha \rho C_\alpha^{\dagger} \right]}{\Tr \left[ C_\alpha \rho C_\alpha^{\dagger} \right]} = \frac{\Tr_{\mcB} \left[ C_\alpha \rho C_\alpha^{\dagger} \right]}{p_\alpha}.
\end{split}\end{align}
In particular, the conditional global states are
$\rho_\alpha  = C_\alpha \rho C_\alpha^{\dagger}/p_\alpha$.

Importantly, the decoherence functional is independent of $t$ (by the cyclic property of the trace) even though the density matrix and class operators are not.  In this sense consistency is a (non-dynamical) property of a state $\rho$, a Hamiltonian, and a set of histories $\{ C_\alpha \}$.





\subsection{Records and accessibility}
\label{subsec:condstates}

The consistency condition, Eq.~\eqref{eq:meddecoh}, is the mixed-state generalization of the requirement that the conditional pure states do not overlap, $\braket{\psi_\alpha}{\psi_\beta} = 0$, and it ensures that Eq.~\eqref{eq:probsum} holds.  For a pure state, the branches of a consistent set of histories will form the tree structure in Figure \ref{fig:purestatepartialorder} and there will necessarily exists \emph{records}, i.e., \ a complete set of orthogonal projectors $R_\alpha$ such that $R_\alpha \ket{\psi} = C_\alpha  \ket{\psi} = \ket{\psi_\alpha}$ \cite{Gell-Mann1993}.  (Note that the class operators $C_\alpha $ are generally not projectors.) 

In the case of a mixed state, the existence of records is formally defined as\footnote{Equation \eqref{eq:mixedrecords} was once called ``strong decoherence'' \cite{Gell-Mann1993}, although that term was later re-appropriated \cite{Gell-Mann1998} for a modified version of Eq.~\eqref{eq:meddecoh}.} \cite{Gell-Mann1993}
\begin{align}
\label{eq:mixedrecords}
R_\alpha \rho = C_\alpha  \rho
\end{align}
for some complete set of orthogonal projection operators $R_\alpha$. This is equivalent to saying that the supports of the conditional states $\rho_\alpha$ lie in orthogonal subspaces.\footnote{We note that this parallels the requirement for the accessibility of the mixed states, i.e., for the states about which one can repeatedly extract information \cite{zurek2013wave-packet}. This is not a coincidence, but we leave a more detailed exploration of this theme for the sequel to the current paper~\cite{riedel2013objective2}.\label{foot9}}  The orthogonality is a sufficient but not necessary condition for Eq.~\eqref{eq:meddecoh} to hold.\footnote{Note that for a given set of histories, it is true that consistency \eqref{eq:meddecoh} with respect to each of pure states $\ket{\psi_i}$ would imply consistency with respect to their statistical mixture $\rho = \sum_i p_i \projector{\psi_i}$, and that consistency with respect to a given $\ket{\psi_i}$ implies the existence of formal records $R^{(i)}_\alpha$.  However, for each $\alpha$ the $R^{(i)}_\alpha$  may differ, so there would not need to be a record $R_\alpha$ for $\rho$ as a whole.}

Records are important because in principle they allow observers within the Universe to become correlated with the history by making a projective measurement.  However, these need not be feasible measurements because the records may not be confined to any particular subsystem; rather, they will usually be encoded across the entire global state \cite{Gell-Mann1993, Gell-Mann1998}. We will emphasize this by calling Eq.~\eqref{eq:mixedrecords} \emph{formal} records.\footnote{We note that Gell-Mann and Hartle called these  \emph{generalized} records \cite{Gell-Mann1993, Gell-Mann1998}.}.   Naively, it might lead us to think that deducing the path taken by a dust grain could require measuring an observable that involves \emph{all} the photons in the room.  The practical accessibility of such records is not guaranteed due to the fact that only consistency of the set of histories has been assumed. In contrast, we aim in this article to describe why reasonably accessible and redundant records of histories should exists in real-world quantum systems (see Ref.~\cite{zwolak2013complementarity} for related discussion).

\subsection{Pure decoherence and approximate consistency}
\label{subsec:consisdecoh}

The decoherence functional satisfies
\begin{align}
| \DD( \alpha, \beta ) |^2 \le \DD( \alpha, \alpha ) \DD( \beta, \beta )
\end{align}
with the probability sum rules obeyed when $| \DD( \alpha, \beta ) |^2 = 0$ for $\alpha \neq \beta$ \cite{Dowker1992}. For realistic quantum systems with simple, physically meaningful projectors, a set of histories will almost never be precisely consistent.  Suppose that the off-diagonal terms of the decoherence functional are small, to some order  $\epsilon$, compared to the diagonal terms:
\begin{align}
| \DD( \alpha, \beta ) |^2 < \epsilon \, \DD( \alpha, \alpha ) \DD( \beta, \beta ), \qquad \alpha \neq \beta.
\end{align}
One can then show that the probability sum rules are valid to the same order $\epsilon$ for almost every possible coarse-graining \cite{Dowker1992}.

One can define, then, for any two different histories $\alpha$ and $\beta$, a consistency factor
\begin{align}\begin{split}
\label{eq:consisfac}
\CF_{\alpha \beta} &= \frac{\DD( \alpha, \beta )}{\sqrt{  \DD( \alpha, \alpha ) \DD( \beta, \beta )}} = \frac{\DD( \alpha, \beta )}{\sqrt{p_\alpha p_\beta }}.
\end{split}\end{align}
We say that the histories are approximately consistent to order $\epsilon$ when $\abs{\CF_{\alpha \beta}} < \epsilon$ for $\alpha \neq \beta$ \cite{McElwaine1996, Dowker1992}.  

The $N(N-1)/2$ consistency factors for a set of $N$ histories are closely related to the $N(N-1)/2$ decoherence factors in the case of pure decoherence of an $N$-dimensional system system. [In particular, see Eq.\ \eqref{eq:consisdecohfactorequiv} below.]  That said, we emphasize that decoherence is a dynamical physical process predicated on a distinction between system and environment, whereas consistency is a timeless property of a set of histories, a Hamiltonian, and an initial state. For a given decohering quantum system, there is generally, but not always, a preferred basis of pointer states \cite{Zurek1981, Zurek1982}.  In contrast, the mere requirement of consistency does not distinguish a preferred set of histories which describe classical behavior from any of the many sets with no physical interpretation. For an arbitrary set of histories, consistency factors will be defined but there may be no system-environment decomposition, and hence no pointer basis or decoherence factors.   Therefore, we will retain the semantic distinction between consistency factor and decoherence factor, using the latter only when there is at least an approximate pointer basis (such as Gaussian wavepackets \cite{Kubler1973, Zurek1993a} or hydrodynamical variables \cite{Gell-Mann1993, Gell-Mann2007, Halliwell1998}).  For more discussion of the connection between the consistency of histories and the physical process of decoherence see Refs.\ \cite{Halliwell1989, ZurekTime, Paz1993, Zurek1993b, Finkelstein1993}

\section{Partial-trace consistency}
\label{sec:rptconsis}          %

We now define the concept of partial-trace consistency \cite{Finkelstein1993}, which bridges the gap between decoherence and consistent histories.   Given a decomposition of Hilbert space into two parts, $\Glob = \mcA \otimes \mcB$, we define a partial-trace decoherence functional by tracing over only $\mcB$:
\begin{align}\begin{split}
\DD_\mcB (\alpha, \beta) &= \Tr_\mcB \left[ C_{\alpha} \rho \, C_{\beta}^{\dag} \right] \\
&= \Tr_\mcB \big[ U_{t_M,t} P_{a_M}^{(M)} U_{t_{M-1},t_M}  \cdots P_{a_1}^{(1)} U_{t,t_1} \\
& \qquad \times \rho \, U_{t,t_1}^\dag P_{b_1}^{(1)} \cdots U_{t_{M-1},t_M}^\dag P_{b_M}^{(M)} U_{t_M,t}^\dag \big]
\end{split}\end{align}
Note that $\DD_\mcB$ is an operator acting on $\mcA$ and has nontrivial dependence on $t$, unlike $\DD$. As we shall see, this reflects the fact that records exist at certain times in certain places.  In contrast to Ref.\ \cite{Finkelstein1993} [see Eq.~(1) therein], we have retained the traditional definition of the class operators, Eq.~\eqref{eq:classop}, using time-evolved projectors.  This allows us to consider a partial-trace decoherence functional evaluated at any time $t$ rather than just at the final time step $t_M$.

We define \emph{partial-trace consistency} with respect to $\mcB$ at some time $t$, or $\mcB$-\emph{consistency}, as
\begin{align}
\DD_\mcB(\alpha, \beta) = 0, \qquad \alpha \neq \beta.
\end{align}
The partial-trace decoherence functional obeys the following basic operator relations:
\begin{gather}
\Tr_\mcA \DD_\mcB(\alpha, \beta) =  \DD(\alpha, \beta), \\
\DD_\mcB(\beta, \alpha) =  \DD_\mcB^\dag(\alpha, \beta), \\
\sum_{\alpha, \beta} \DD_\mcB(\alpha, \beta) = \Tr_\mcB \rho  = \rho^\mcA .
\end{gather}

When originally introduced, the partial-trace decoherence functional was taken with respect to the environment ($\mcA \to \Sys$, $\mcB \to \Env$).  For our purposes, we will assume that the environment can be broken into a fragment and its complement, $\Env = \Frag \otimes \FragBar$, and study $\Frag$-consistency.  

We note that in quantum Darwinism the fragment of the environment that is traced over is usually $\FragBar$, as the quantity of interest is the mutual information between $\Sys$ and $\Frag$, which depends on their reduced state (i.e., without $\FragBar$). For $\Frag$-consistency, the trace will instead be taken over $\Frag$. In both cases, however, one is ultimately interested in the record deposited in $\Frag$ (a small fragment of the whole environment). Quantum Darwinism requires that $\Frag$ contains a record, which will ensure $\Frag$-consistency. Of course, the requirement of consistency is not strict enough on its own to guarantee the presence of a record for the case of initially mixed states, as we have emphasized both above and below.

\subsection{Properties}
\label{subsec:ptproperties}

We now collect some known properties of the partial-trace decoherence functional.  We fill in those proofs that were originally absent from Ref.\ \cite{Finkelstein1993} and make the extension from $\Env$-consistency to $\Frag$-consistency where appropriate.  We will occasionally be sloppy with the notation for the tensor product decomposition by interchanging $\Sys \otimes \Frag \otimes \FragBar$ and  $\Sys \otimes \FragBar \otimes \Frag$, but the meaning will be clear.

\begin{description}

\item[Inheritance of consistency] \label{it:metaconsis}  \hfill \\
Any histories which are $\Frag$-consistent at time $t$ are automatically $\Env$-consistent and therefore consistent.
\begin{align}\begin{split}
\DD_\Frag(\alpha, \beta) = 0 &\quad \Rightarrow \quad \DD_\Env(\alpha, \beta) = 0 \\
&\quad \Rightarrow \quad \DD(\alpha, \beta) = 0.
\end{split}\end{align}
This follows immediately from taking the trace.

\item[Consistency implies diagonalization] \hfill \\
For trajectories of the system $\Sys$ (i.e., histories constructed of projectors of the form $P = P^\Sys \otimes I^\Env$), $\Env$-consistency (and hence $\Frag$-consistency) at the final time step $t = t_M$ implies the density matrix of the system $\rho^\Sys$ is block diagonal on subspaces associated with those projectors.


To show this, sum $\DD_\Env(\alpha, \beta)$  over all projectors for $\alpha$ and $\beta$ except at the final time.  If $a_M \neq b_M$, then $\alpha \neq \beta$ and
\begin{align}\begin{split}
0 &= \sum_{a_1} \cdots \sum_{a_{M-1}} \sum_{b_1} \cdots \sum_{b_{M-1}} \left[ \DD_\Env(\alpha, \beta) \vert_{t_M} \right] \\
&= \Tr_\Env \left[ (P^{\Sys}_{a_M} \otimes I^\Env) \left( \rho \, \vert_{t_M} \right) \, (P^{\Sys}_{b_M} \otimes I^\Env) \right] \\
&= P^{\Sys}_{a_M} \left( \rho^\Sys \, \vert_{t_M} \right)  P^{\Sys}_{b_M}.
\end{split}\end{align}

In fact, $\Frag$-consistency implies that off-diagonal blocks of $\rho^{\Sys \FragBar}$ at the final time $t_M$ vanish,
\begin{align}\begin{split}
0 &= \sum_{a_1} \cdots \sum_{a_{M-1}} \sum_{b_1} \cdots \sum_{b_{M-1}} \left[ \DD_\Frag(\alpha, \beta) \vert_{t_M} \right] \\
&= \left( P^{\Sys}_{a_M} \otimes I^\FragBar \right) \left( \rho^{\Sys \FragBar} \, \vert_{t_M} \right) \left( P^{\Sys}_{b_M} \otimes I^\FragBar \right),
\end{split}\end{align}
so the total state of $\Sys \FragBar$ is block diagonal:
\begin{align}
\qquad \rho^{\Sys \FragBar} \, \vert_{t_M} = \sum_{a_M}  \left(P^{\Sys}_{a_M} \otimes I^\FragBar \right) \left( \rho^{\Sys \FragBar} \, \vert_{t_M} \right)  \left( P^{\Sys}_{a_M} \otimes I^\FragBar \right).
\end{align}
This is equivalent to having zero quantum discord \cite{Zurek2000,Ollivier2001,Henderson2001} from $\Sys$ to $\FragBar$ at the final time (see Ref. \cite{zwolak2013complementarity} for a discussion of discord in the context of quantum Darwinism). 

\item[Generalized sum rule for probabilities]\label{it:sysconsiscond} \hfill\\
$\Env$-consistency implies that the conditional states of the system, $\rho^\Sys_\alpha  = \Tr_\Env \left[C_\alpha \rho \, C_\alpha^\dag  \right]/p_\alpha$, obey
\begin{align}
\label{eq:bieffconsiscondition}
p_{\alpha \vee \beta} & \rho^\Sys_{\alpha \vee \beta}
= p_\alpha \rho^\Sys_{\alpha} + p_\beta \rho^\Sys_{ \beta} .
\end{align}
This is a generalization of the probability sum rule, Eq.~\eqref{eq:probsum}.  To prove Eq.~\eqref{eq:bieffconsiscondition}, recall that $C_{\alpha \vee \beta} = C_{\alpha} + C_\beta$ so
\begin{align}\begin{split}
\quad \DD_\Env (\alpha \vee \beta,\alpha \vee \beta)
&= \DD_\Env(\alpha ,\alpha  ) + \DD_\Env(\alpha ,  \beta) \\
& \qquad + \DD_\Env(  \beta,\alpha  ) + \DD_\Env(  \beta,  \beta) \\
&= \DD_\Env(\alpha ,\alpha  ) + \DD_\Env(  \beta,  \beta) ,
\end{split}\end{align}
where the second line is due to $\Env$-consistency.  We get Eq.~\eqref{eq:bieffconsiscondition} from the definition of the conditional state and the fact that $p_{\alpha \vee \beta} = p_\alpha + p_\beta$ by the consistency of the histories.

\item[Extension of histories] \hfill \\
Insofar as the fragment decouples from the system and the rest of the environment -- thus precluding recoherence -- any two particular histories $\alpha$ and $\beta$ that are $\Frag$-consistent will remain so when extended into the future with additional histories $\alpha'$ and $\beta'$ of $\Sys\FragBar$.  More precisely, suppose that $\DD_\Frag(\alpha, \beta) \, \vert_{t_*} = \Tr_\Frag [C_\alpha \rho \, C_\beta^\dag ] \, \vert_{t_*} =0$ at some fixed time $t_* \ge t_M$, and that the Hamiltonian for times after $t_*$ takes the form $\hat{H}' = \hat{H}^{\Sys \FragBar} \otimes I^\Frag + I^{\Sys \FragBar} \otimes \hat{H}^\Frag$.  Suppose also that the extended history $\alpha \wedge \alpha'$ running to some later time $t_{M+M'} > t_*$ is of the form $C_{\alpha \wedge \alpha'} = C_{\alpha'} C_{\alpha}$, where\footnote{In particular, \eqref{eq:extended-class-op} applies when $\alpha$ and $\alpha'$ are histories concerning only $\Sys$.}
\begin{align}\begin{split}
\label{eq:extended-class-op}
C_{\alpha'} &= C_{\alpha'}^{\Sys \FragBar} \otimes I^\Frag\\
&=  P_{{a}_{M'}^\prime}^{(M+M')\Sys \FragBar }(t_{M+M'}) \otimes I^\Frag \\
& \qquad \times \cdots \times  P_{{a}_{1}^\prime}^{(M+1)\Sys \FragBar }(t_{M+1}) \otimes I^\Frag, 
\end{split}\end{align}
with $t_{M} < t_{M+1} < \cdots < t_{M+M'}$. If a similar extension is made for the history $\beta$, then the two extended histories can be seen to be $\Frag$-consistent for all times\footnote{Note that when $t < t_*$, \eqref{eq:decoupled-fragment} need not hold because we are not guaranteed that $U_{t_*,t} = U_{t_*,t}^{\Sys \FragBar } \otimes U_{t_*,t}^{\Frag }$.} $t \ge t_*$:
\begin{align}\begin{split}
\label{eq:decoupled-fragment}
\DD_\Frag &(\alpha \wedge \alpha', \beta \wedge \beta') \\
&= \Tr_\Frag \left[C_{\alpha \wedge \alpha'} \rho \, C_{\beta \wedge \beta'}^\dag \right] \\
&= C_{\alpha'}^{\Sys \FragBar}   \Tr_\Frag \left[ C_{\alpha} \rho \, C_{\beta}^\dag  \right]  C_{\beta'}^{\Sys \FragBar \dag} \\
&= C_{\alpha'}^{\Sys \FragBar} U_{t_*,t}^{\Sys \FragBar } \big(\DD_\Frag(\alpha , \beta ) \, \vert_{t_*} \big) U_{t_*,t}^{\Sys \FragBar \dag} C_{\beta'}^{\Sys \FragBar \dag} \\
&= 0
\end{split}\end{align}

The ability of small decoupled fragments to guarantee partial-trace consistency was discussed in Ref.\ \cite{Gell-Mann1998}.  It is important because partial-trace consistency is a time-dependent statement, so it is not obvious \emph{a priori} that even trivial extensions of histories in time are possible.  (Indeed, if $\Frag$ remains coupled to the rest of the environment, then $\Frag$-consistency can be destroyed in time even when the histories are unchanged.)  For $\alpha = \beta$, the question of whether $\alpha \wedge \alpha'$ is consistent with $\beta \wedge \beta'$ of course depends on the details of the extensions $\alpha'$ and $\beta'$, but this lemma shows that, at the least, two distinct $\Frag$-consistent histories $\alpha$ and $\beta$ remain $\Frag$-consistent when extended to cover future events so long as $\Frag$ is undisturbed.

\end{description}

As emphasized by Finkelstein, these properties are evidence that $\Env$-consistency is the natural requirement for saying that a set of histories are consistent because of the physical process of decoherence, i.e., that the histories truly \emph{decohere} rather than merely attain consistency.  The inherent time dependence of the partial-trace decoherence functional emphasizes the important distinction between the physical process of decoherence (which occurs between two quantum systems, such as $\Sys$ and $\Frag$, during an interval of time) and the mathematical condition of consistency (which applies timelessly to a set of histories, and which does not require a decomposition of the Hilbert space into subsystems). Finkelstein's partial-trace consistency elegantly links these two concepts.

\subsection{Partial-trace consistency factor}
\label{subsec:parconsisdecoh}

For an arbitrary decomposition $\Glob = \mcA \otimes \mcB$, the partial-trace decoherence functional has matrix elements bounded as
\begin{align}
\label{eq:ptbound}
\abs{\matrixelement{A_i}{\DD_\mcB(\alpha, \beta)}{A_j}}^2 \le \matrixelement{A_i}{\DD_\mcB(\alpha, \alpha)}{A_i}\matrixelement{A_j}{\DD_\mcB(\beta, \beta)}{A_j}
\end{align}
for any basis $\ket{A_i}$ of $\mcA$ \cite{Finkelstein1993}.  This suggests that the proper way to generalize the approximate consistency condition for the partial-trace decoherence functional is to require that the left-hand side of Eq.~\eqref{eq:ptbound} be much smaller than the right-hand side for all $i$ and $j$ \cite{Finkelstein1993}.  More specifically, define a partial-trace consistency factor
\begin{align}\begin{split}
\CF^\mcB_{\alpha,\beta} &= \frac{\DD_\mcB(\alpha, \beta)}{\sqrt{\Tr \DD_\mcB(\alpha, \alpha) \Tr \DD_\mcB(\beta, \beta)}} \\
&= \frac{\Tr_\mcB\left[ C_\alpha \rho C_\beta^\dag \right]}{\sqrt{ p_\alpha p_\beta}}.
\end{split}\end{align}
We can then say the histories are approximately partial-trace consistent when the operator $\CF^\mcB_{\alpha,\beta}$ is small according to an appropriate norm.

We now relate the partial-trace consistency factor to the decoherence factor and to the fidelity between conditional states.

\begin{description}

\item[Decoherence factor and consistency factor]\hfill \\
Consider the case of pure decoherence \eqref{eq:puredecoh} discussed in Section \ref{subsec:puredecoh} with a simple set of histories
\begin{align}
\label{eq:puredecohhistory}
\{ C_\alpha =  P_a = \projector{a} \otimes I^\Env \},
\end{align}
where the projectors act non-trivially only on $\Sys$ and pick one of the pointer states $\ket{a}$.  
The partial-trace consistency factor for the environment $\Env$ is
\begin{align}\begin{split}
\label{eq:consisdecohfactorequiv}
\CF^\mathcal{\Env}_{\alpha,\beta} &= \Gamma_{a b}  \ket{a} \bra{b} .
\end{split}\end{align}
When (as considered in Section \ref{subsec:fragmentation}) there are no intra-environmental interactions and there is a product initial state, we can more generally write
\begin{align}\begin{split}
\CF^\mathcal{\Frag}_{\alpha,\beta} &=\Gamma^\Frag_{ ab}  \ket{a} \bra{b} \otimes U_a^\FragBar \rho^\FragBar U_b^{\FragBar}\vphantom{U}^\dag .
\end{split}\end{align}
The operator $U_a^{\FragBar} \rho^\FragBar U_b^{\FragBar} \vphantom{U}^\dagger$ need not be a density matrix, but for any unitarily invariant norm $||\cdot||$ (like the trace norm; see below) we have $||\CF^\mathcal{\Frag}_{\alpha,\beta}|| = \Gamma^\Frag_{a b} ||\rho^{\FragBar,0}||$, where $||\rho^{\FragBar,0}||$ is fixed by initial conditions.  Thus, for pure decoherence without intra-environmental interactions, the decoherence factor $\Gamma^\Frag_{a b}$ controls the norm of the consistency factor $\CF^\mathcal{\Frag}_{\alpha,\beta}$.  One  vanishes if and only if the other does.

\item[Fidelity and the consistency factor] \hfill \\
When the global state is pure, $\rho = \projector{\psi}$, the trace norm of the partial-trace consistency factor equals the fidelity between the relevant conditional states:
\begin{align}
\label{eq:fidelity-consistency}
F(\rho^\mcA_\alpha,\rho^\mcA_\beta) = \AAbs{\CF^\mcA_{\alpha,\beta}}_1,
\end{align}
where $F(\sigma_1,\sigma_2) = \AAbs{\sqrt{\sigma_1}\sqrt{\sigma_2}}_1$ is the quantum fidelity and $\aabs{W}_1 = \Tr \abs{W} = \Tr \sqrt{W^\dagger W}$ is the trace norm\footnote{Note that we use the un-squared convention; some authors refer to $F^2$ as the fidelity.} of an operator $W$.

To show this, let the Schmidt decomposition of the branches $\ket{\psi_\alpha}$ be
\begin{align}
\ket{\psi_\alpha} = \sqrt{p_\alpha} \sum_i \sqrt{d_i^{(\alpha)}} \ket{A_i^{(\alpha)}}\ket{B_i^{(\alpha)}}
\end{align}
where, for each $\alpha$, the $d_i^{(\alpha)}$ are positive coefficients and the $\ket{A_i^{(\alpha)}}$ and $\ket{B_i^{(\alpha)}}$ form orthonormal bases for $\mcA$ and $\mcB$.  Then the partial-trace consistency factors are
\begin{align}\begin{split}
\qquad \CF^\mcA_{\alpha,\beta} &= \sum_{i,j} \sqrt{d_i^{(\alpha)} d_j^{(\beta)}} \braket{A_j^{(\beta)}}{A_i^{(\alpha)}} \ketbra{B_i^{(\alpha)}}{B_j^{(\beta)}}.
\end{split}\end{align}
On the other hand, the conditional states of $\mcA$ are
\begin{align}
\quad \rho^\mcA_\alpha = \frac{\Tr_\mcB \left[ \projector{\psi_\alpha} \right]}{p_\alpha} = \sum_i d_i^{(\alpha)} \projector{A_i^{(\alpha)}}.
\end{align}
Direct computation then shows that
\begin{align}
F(\rho^\mcA_\alpha,\rho^\mcA_\beta)  = \Tr \sqrt{M} = \AAbs{\CF^\mcA_{\alpha,\beta}}_1
\end{align}
where $M$ is a matrix with elements
\begin{align}\begin{split}
M_{i,i'}= \sum_{j}& \braket{A_i^{(\alpha)}}{A_j^{(\beta)}} \braket{A_j^{(\beta)}}{A_{i'}^{(\alpha)}}  \\
&\times d_j^{(\beta)} \sqrt{d_i^{(\alpha)} d_{i'}^{(\alpha)}}.
\end{split}\end{align}

\end{description}

\subsection{Records and consistency}
\label{subsec:recordsforptc}

We now discuss the relationship between records and consistency.

\begin{description}

\item[Conditions for records] \label{it:record} \hfill \\
The following statements are equivalent conditions for saying a record of the history $\alpha$ exists in $\mcB$.

\begin{enumerate}

   \item The conditional states  $\rho_\alpha^\mcB = \Tr_\mcA [ \projector{\psi_\alpha} ]/p_\alpha$ are orthogonal. That is, their respective supports are mutually orthogonal subspaces.\footnote{See footnote \ref{foot9}.}

   	\item The fidelities of the conditional states obey $F(\rho^\mcB_\alpha, \rho^\mcB_\beta) = \delta_{\alpha, \beta}$.

   \item \label{it:recordc}There are orthogonal projectors $R_\alpha = I^{\mcA} \otimes R_\alpha^\mcB$ acting nontrivially only on $\mcB$ which pick out the conditional states $\rho_\alpha$:
\begin{align}
\label{eq:record-block}
\left( I^{\mcA} \otimes R_\alpha^\mcB  \right) \rho= C_\alpha \rho,
\end{align}
which implies $R_{\alpha} \rho  R_{\alpha} = C_\alpha \rho C_\alpha^\dag = p_\alpha \rho_\alpha$.
\end{enumerate}
This means that an observer could realistically determine the history $\alpha$ by making only a local measurement on $\mcB$.


That (1) $\Leftrightarrow$ (2) is a basic property of the quantum fidelity \cite{NielsenText}.  We get (1) $\Leftrightarrow$ (3) by defining $R_\alpha^\mcB$ to project onto the support of $\rho_\alpha^\mcB$ and expanding with the Schmidt decomposition
\begin{align}
\label{eq:schmidt-record}
C_\alpha \ket{z} = \sum_{j} \gamma_{j,z}^{(\alpha)} \ket{A_{j,z}^{(\alpha)}}\ket{B_{j,z}^{(\alpha)}}.
\end{align}
of the state $C_\alpha \ket{z}$ over the $\mcA$-$\mcB$ division, where $\rho = \sum_z \lambda_z \projector{z}$ (and $\lambda_z > 0$).


\item [Records imply consistency] \hfill \\
When there is a record of the history in $\mcB$, the histories are $\mcB$-consistent.

This can be shown by expanding the condition for orthogonality of the conditional states ($\Tr_\mcB [\rho_\alpha^\mcA \rho_\beta^\mcA]=0$ when $\alpha \neq \beta$) with Eq.~\eqref{eq:schmidt-record} to see that $\braket{B_{j,z}^{(\alpha)}}{B_{j',z'}^{(\beta)}}=0$ for all $j,j',z,z'$ whenever $\alpha \neq \beta$.  $\mcB$-consistency then follows by evaluating $\DD_\mcB (\alpha,\beta)$ with Eq.~\eqref{eq:schmidt-record}.



\item [Consistency with purity implies records] \hfill \\
When the global state is pure, $\rho = \projector{\psi}$,  $\mcB$-consistency implies there is a record of the history in $\mcB$.  This follows from Eq.~\eqref{eq:fidelity-consistency}.

\end{description}

In short: the existence of a local record is a sufficient condition for partial-trace consistency.  When the global state is pure, it is a necessary condition.  This is an extension of the previously known relationship between formal records and consistency (in the consistent histories setting) and between records and decoherence.

\subsection{Records correlated in time}
\label{subsec:recordsintime}

Consider an $\Frag$-consistent set of histories $\{ \alpha \}$ of the system $\Sys$ with a pure global state $\rho = \projector{\psi}$, so the tree structure in Figure \ref{fig:purestatepartialorder} is realized.  By the last property in the previous section, we know that there exist branches $\ket{\psi_\alpha } \in \Sys \otimes \FragBar \otimes \Frag_\alpha $, where the $\Frag_\alpha$ are mutually orthogonal subspaces of $\Frag$.  Now, let $\{ \hat{\alpha} \}$ be a coarse-grained set of histories obtained by summing over the projectors at the final time step:
\begin{align}\begin{split}
C_{\hat{\alpha}} &= \sum_{a_M} C_\alpha \\
&= \sum_{a_M} P_{a_M}^{(M)}(t_M) P_{a_{M-1}}^{(M-1)}(t_{M-1}) \cdots P_{a_1}^{(1)}(t_1) \\
&=  P_{a_{M-1}}^{(M-1)}(t_{M-1}) \cdots P_{a_1}^{(1)}(t_1) .
\end{split}\end{align}
Then
\begin{align}
\ket{\psi_{\hat{\alpha}} } = \sum_{a_M} \ket{\psi_\alpha } \in \Sys \otimes \FragBar \otimes \Frag_{\hat{\alpha}}
\end{align}
where the $\Frag_{\hat{\alpha}} = \bigoplus_{a_m} \Frag_\alpha$ are direct sums of orthogonal subspaces $\Frag_\alpha$.  The $\Frag_{\hat{\alpha}}$ are therefore also orthogonal.

By repeatedly summing over the final set of projectors, we can continue this process, showing that a branch structure on the Hilbert space of $\Frag$ can be found that is equivalent to the branch structure of the $\ket{\psi_\alpha}$ (seen in Figure \ref{fig:purestatepartialorder}).  More precisely, it follows that for all $\bar{m}$ (with $1 \le \bar{m} \le M$) we can decompose
\begin{align}
\Frag_{(a_1, \ldots, a_{\bar{m}-1})} = \bigoplus_{a_{\bar{m}}}  \Frag_{(a_1, \ldots, a_{\bar{m}-1}, a_{\bar{m}})}
\end{align}
into mutually orthogonal subspaces where
\begin{align}
\ket{\psi_{(a_1, \ldots, a_{\bar{m}})} } \in \Sys \otimes \FragBar \otimes \Frag_{(a_1, \ldots, a_{\bar{m}})}.
\end{align}

If we allow for a non-interacting auxiliary environment $\Aux$ to be appended\footnote{The inclusion of $\Aux$ ensures that there will be sufficient dimensionality in $\tilde{\Frag}$ to fit the entire tensor product, Eq.~\eqref{eq:fragtime}.  This might not otherwise be true if some histories have zero probability.} to the fragment $\Frag$, then there exists a decomposition of the supplemental fragment $\tilde{\Frag} = \Frag \otimes \Aux$ into subsystems
\begin{align}
\label{eq:fragtime}
\tilde{\Frag} = \tilde{\Frag}^{(1)} \otimes \dots \otimes \tilde{\Frag}^{(M)}
\end{align}
and each subsystem into subspaces
\begin{align}
\tilde{\Frag}^{(m)} = \bigoplus_{a_m} \tilde{\Frag}^{(m)} _{a_m}
\end{align}
(for each $m$) such that
\begin{align}\begin{split}
\ket{\psi_{(a_1, \ldots, a_M)}} &\in \Sys \otimes \FragBar \otimes \Frag_{(a_1, \ldots, a_M)} \\
&\subset \Sys \otimes \FragBar \otimes \tilde{\Frag}^{(1)} _{a_1} \otimes \dots \otimes \tilde{\Frag}^{(M)} _{a_M}.
\end{split}\end{align}
In other words, an observer can infer the state of $\Sys$ at time $t_m$ [more precisely, which of the projectors $P_{a_m}^{(m)}$ it satisfied] by making a measurement of only subsystem $\tilde{\Frag}^{(m)}$ at later time $t \ge t_M > t_m$.

We emphasize that this subsystem structure is only guaranteed by $\Frag$-consistency when the global state is pure. Of course, there is no need for the decomposition in Eq.~\eqref{eq:fragtime} to have anything to do with a natural separation of the environment into parts accessible to a realistic observer, so such a measurement may be completely infeasible.  In a forthcoming paper \cite{Riedel01}, though, we show that collisional decoherence by photons provides a natural example where this decomposition is physically meaningful and exploited every day.

\section{CNOT example}
\label{sec:CNOT-example}

In this section we apply the concepts introduced above to an example of a two-state system $\Sys$ flipping back and forth while being intermittently measured by a large multi-partite environment $\Env$ through controlled-not interactions \cite{zurek2003a, NielsenText}.  The dynamics are not meant to closely resemble a physical system, but merely to serve as a minimal example on which we can demonstrate the mathematical machinery.

Let $\Sys$ be spanned by eigenstates $\ket{0}_\Sys$ and $\ket{1}_\Sys$ (which will turn out to be pointer states), and let the total environment $\Env = \bigotimes_{m=1}^M \Env_m = \bigotimes_{k=1}^{\Es} \varepsilon_k$ be separated into $M$ sub-environments $\Env_m$, each of which is composed of many identical two-state subsystems $\varepsilon_k$ spanned by $\ket{0}_{k}$ and $\ket{1}_{k}$. Here,  $k=1,\ldots,\Es$ runs over all the components of $\Env$.


%



\begin{figure}
  \centering
  \includegraphics[width=8cm]{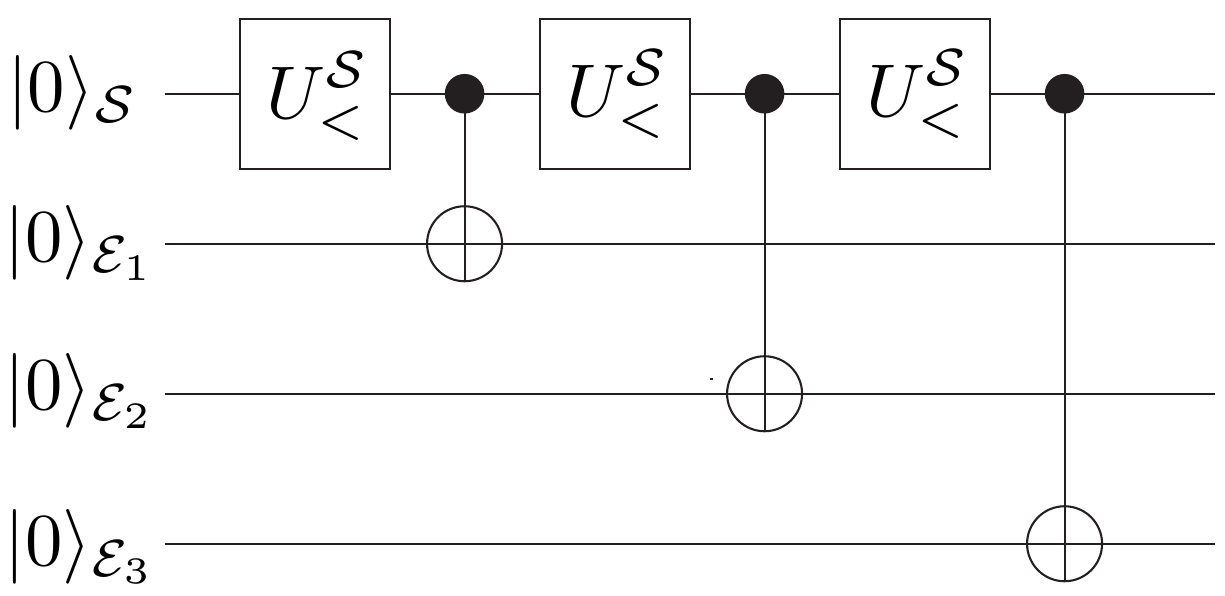}
  \caption{Evolution in the \CNOT{} example. The system starts out in the state $\ket{0}_\Sys$ and then undergoes three discrete branching and recording events. During each recording event, the system is perfectly decohered in the $\ket{0}$, $\ket{1}$ basis. The environment that acquires the record at each event $m$, $\Env_m$, will be composed of many spins (even though it is shown for simplicity as a single spin), and thus redundant records are generated.}
  \label{fig:cnotcirc}
\end{figure}

\begin{figure*} [t]
  \centering
  \includegraphics[width=16cm]{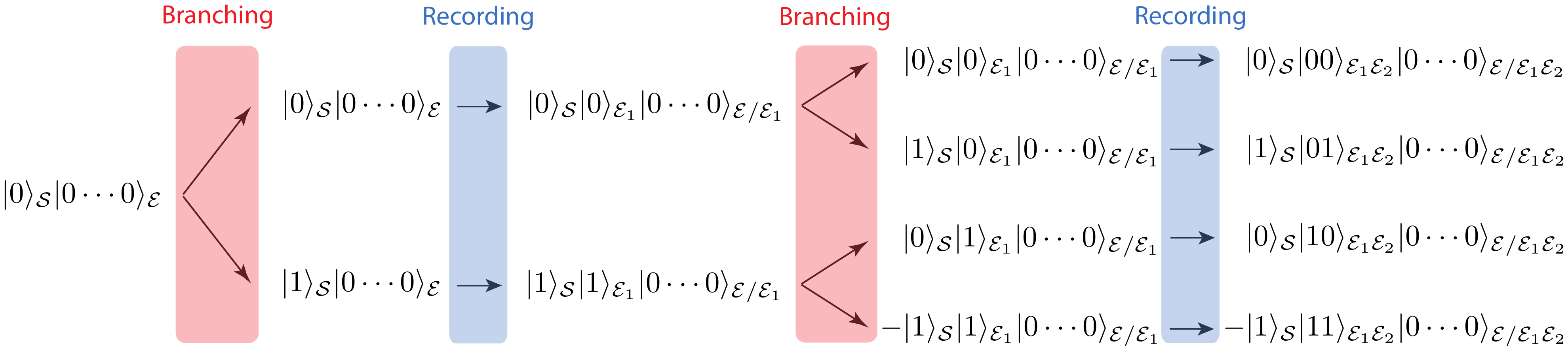}
  \caption{Branching structure of the evolution in the \CNOT{} example. An initial product state of the system and all the environment subsystems evolves according to sequential branching and recording events. Here, the records are into a single two level subsystem of the environment for ease of depiction. However, each recording event can be redundant, making many copies of the system's state after each branching event.}
  \label{fig:cnot}
\end{figure*}

We consider the unitary governing the discrete self-evolution of the system to be
\begin{align}
U_{\branch}^\Sys = \frac{1}{\sqrt{2}}\left( \begin{array}{cc}
1 & 1 \\
1 & -1 \end{array} \right), 
\end{align}
where ``$\branch$'' signifies a branching event. (This is a Hadamard transformation.)
This has the effect of causing an initial pointer state to rotate into a balanced superposition:
\begin{align}
U_{\branch}^\Sys \ket{0}_\Sys &= \frac{\ket{0}_\Sys + \ket{1}_\Sys}{\sqrt{2}},\\
U_{\branch}^\Sys \ket{1}_\Sys &= \frac{\ket{0}_\Sys - \ket{1}_\Sys}{\sqrt{2}}.
\end{align}
The intermittent \CNOT{} interactions are described by Eq.~\ref{eq:CSHIFT} with
\begin{align}
U^{\Sys \varepsilon_k}_{\CN} = \ket{0}_\Sys \bra{0} \otimes I^{\varepsilon_k} + \ket{1}_\Sys \bra{1} \otimes \big(\ket{0}_{k} \bra{1} +\ket{1}_{k} \bra{0} \big).
\end{align}
When the spins in the subenvironment $\Env_m$ each record the pointer state of the system, then the interaction is described by
\begin{align}
U_{\CN}^{\Sys \Env_m} = \prod_{\varepsilon_k \in \Env_m} U_{\CN}^{\Sys k}.
\end{align}
An initially unentangled state
\begin{align}
\ket{\psi^0} = \ket{0}_\Sys \ket{0}_{1}\cdots \ket{0}_{\Es} \label{eq:InitState}
\end{align}
that branched three times ($M = 3$), and was recorded between branching events, would be described by
\begin{align}\begin{split}
\ket{\psi} = U_{\CN}^{\Sys \Env_3} U_{\branch}^\Sys U_{\CN}^{\Sys \Env_2} U_{\branch}^\Sys U_{\CN}^{\Sys \Env_1} U_{\branch}^\Sys \ket{\psi^0}.
\end{split}\end{align}
This evolution is shown schematically in Figs. \ref{fig:cnotcirc} and \ref{fig:cnot}.
It will produce the global state
\begin{align}\begin{split}
\label{eq:CNOT-branches}
\ket{\psi} &= \sum_\rec \sqrt{p_{\rec}} \ket{a_M}_\Sys \ket{\rec}_\Env \\
&= \sqrt{p_{000}} \ket{0}_\Sys \ket{\alpha=000}_\Env +
\sqrt{p_{001}} \ket{1}_\Sys \ket{\alpha=001}_\Env \\
&\qquad +
\sqrt{p_{010}} \ket{0}_\Sys \ket{\alpha=010}_\Env + \cdots
\end{split}\end{align}
where $\rec$ ranges over the possible histories of the system (i.e., $000, 001, 010, \ldots, 111$), $p_{\rec} = 1/8$ are uniform probabilities [when the initial state is Eq.~(\ref{eq:InitState})], and $\ket{a_M}$ are the pointer states of the system at time $t_M$ (i.e., the last entry in the history $\rec$).  The subenvironment states are redundant records of the history of the system, i.e., when each subenvironment has three spins ($\Es = 3M = 9$), the state $\ket{\alpha=010}_\Env$ would be $\ket{000111000}_\Env$. 
For any fragment $\Frag$ of the environment, the pure  states $\ket{\rec}_{\Frag}$ are well defined because the conditional states $\ket{\rec}_{\Env}$ of the environment are pure product states.  

As is commonly the case, set selection is an issue: There are infinitely many sets of consistent histories that may be chosen for this evolution, which we demonstrate in the Appendix by constructing a wide class of mutually incompatible sets. The branches corresponding to any \emph{redundantly} consistent set must, however, be the ones in Eq.~\eqref{eq:CNOT-branches}, or a sum thereof. (For the purposes of this example, we define ``redundantly'' to mean at least three times; that is, ``redundantly consistent'' means partial-trace consistent for at least three disjoint fragments of the environment.) 

The uniqueness of the branches for a redundantly consistent set can be shown in the following way: Recall that the existence of local records is equivalent to partial-trace consistency since the state is pure.  Any possible partial trace of the global state produces a density matrix with rank at most eight, so no more than eight histories can possibly be locally recorded in a fragment.  The only fragments with density matrices that achieve a rank of eight are those that contain a spin from each of $\Env_1$, $\Env_2$, and $\Env_3$.  (The system $\Sys$ itself can be considered to be a member of $\Env_3$; it is only two-dimensional, and contains a record of its own final state but not its entire history.)  The branches $\ket{a_M}_\Sys\ket{\rec}_{\Env}$ in Eq.~\eqref{eq:CNOT-branches} are clearly recorded redundantly in each such fragment,
and no alternative choice of eight branches can exist by the triorthogonal decomposition theorem\footnote{The triorthogonal decomposition theorem states that if $\ket{\psi}=\sum_i d_i \ket{A_i}\otimes \ket{B_i}\otimes \cdots \otimes \ket{Z_i}$ is the state of at least three subsystems, where the $\ket{A_i}$, $\ket{B_i}$, and so on are sets of orthogonal vectors local to their respective subsystems, then there is no alternative decomposition $\ket{\psi}=\sum_i {d'}_i \ket{{A'}_i} \otimes \ket{{B'}_i} \otimes \cdots \otimes \ket{{Z'}_i}$ unless each alternative set of vectors differs only trivially from the set it replaces \cite{elby1994triorthogonal}.  Note that in order to use this to prove uniqueness of the branches in Eq.~\eqref{eq:CNOT-branches} we must sometimes consider multiple bits in the environment to be one subsystem for the purposes of the theorem.} \cite{elby1994triorthogonal} so long as each sub-environment has three or more spins.  Finally, one can use an extension of this theorem to prove that there is no choice of seven or fewer branches that are recorded in three fragments unless they are each formed as the sum of the $\ket{a_M}_\Sys\ket{\rec}_{\Env}$ \cite{riedel2013local}.

With the branches in hand, the projectors forming the redundantly consistent set can be selected in many ways. The most straightforward choice are the projectors onto the pointer states of the system  at intermediate times. However, it's also possible to obtain the same branch vectors (and hence, the same corresponding probabilities) by using projectors onto the records in the environment. Even for redundant records, there are many such choices, i.e., different subsets of spins of the environment. Thus, the requirement of redundant consistency does not fully determine the histories (some trivial freedoms remain), but it does uniquely fix the branch vectors.

Moreover, this preferred decomposition -- the unique branches corresponding to the redundantly consistent histories -- and the special role of the system qubit relative to the environmental qubits are fully encoded in the evolving global state; it is not put in through a choice of projectors. Rather, the branches are determined by the initial global state $\vert \psi^0 \rangle$, the Hamiltonian $\hat{H}$ governing the dynamics, and the decomposition of Hilbert space into parts -- without distinguishing one of those parts as a preferred system within the consistent histories formalism.

If the environment was initially in a mixed state -- as is the case for photons -- then one can draw similar conclusions about the redundant records of the branches. Let us extend the above example to the case where every environment spin is initialized in the state
\begin{align}
\rho^{k,0} = \left( \begin{array}{cc}
p_0 & 0 \\
0 & p_1 \end{array} \right)
\end{align}
instead of the $\ket{0}$ state, where $p_0$ ($p_1=1-p_0$) is the probability that the environment spin is in the $\ket{0}$ ($\ket{1}$) state. After one branching and recording event, the states
\begin{align}
\rho_0^{\Env_1} = \left( \begin{array}{cc}
p_0 & 0 \\
0 & p_1 \end{array} \right)^{\otimes \Es_1} , \,\, \rho_1^{\Env_1} = \left( \begin{array}{cc}
p_1 & 0 \\
0 & p_0 \end{array} \right)^{\otimes \Es_1}
\end{align}
are the imprints left on the first subenvironment $\Env_1$ by the respective pointer states $\ket{0}_\Sys$, $\ket{1}_\Sys$ of the system.  (This is to be compared to the two orthogonal states $\ket{0}^{\otimes \Es_1}$ and $\ket{1}^{\otimes \Es_1}$ if the environment is initially in the pure state $\ket{0}$.) Clearly, now the two branches do not have a perfect record of the state of the system. Indeed, mixed states and other factors will lead us to examine imperfect records. A brief discussion follows. We will leave a more thorough account of this for later~\cite{riedel2013objective2}. 

For a single spin in the first subenvironment ($\Env_1 = \varepsilon_1$), the orthogonality of the records is $1-F = 1 - 2 \sqrt{p_0 p_1}$ in terms of fidelity. For many spins in $\Env_1$, however, the record becomes nearly perfect (and perfect in the case of an infinite number of environment spins), $1-F = 1- (2 \sqrt{p_0 p_1})^{\Es_1}$. Moreover, for many spins in the environment, the (imperfect) records are redundant. This is a particular example of a purely decohering interaction always giving rise to redundant records~\cite{zwolak2014amplification}. 

To connect with the discussion above, when the environment is mixed but sufficiently large, there are many subsets of the environment that allow for an orthogonal projector $R_\alpha$ to be constructed so  \eqref{eq:record-block} approximately holds. Restricting to histories formed from projectors on the system only, one can potentially choose any set of projectors on the system for each and every time step. Considering projectors other than those for the pointer states of the system will result in lower fidelity records -- potentially eliminating them entirely -- and a reduction in their redundancy~\cite{Ollivier2004,Ollivier2005,zwolak2013complementarity}.

When many branching events occur, the same conclusions hold: The redundant records deposited in the environment are stored in states of the form 
\begin{equation}
	\rho_{\alpha_1}^{\otimes \Es_1} \otimes \rho_{\alpha_2}^{\otimes \Es_2} \otimes \rho_{\alpha_3}^{\otimes \Es_3}  \otimes \cdots
\end{equation}
where $\alpha_m$ is one of the possible pointer outcomes at time $t_m$. Any choice of projectors that select branches which do not correspond to pointer states will be associated with degraded records, just as in the case of the single time history above. Furthermore, by purifying the initial environment states, one can again uniquely fix the branch vectors as above. The accessibility of the records that lie on those branches to local observers -- ones with access only to the unpurified environment -- will be diminished by the mixedness of the environment.

While the simple example above is meant to illustrate the mathematical machinery, repetitive monitoring by distinct parts of the environment mimics the real-world photon environment that we extract most of our information from: As an object moves, it scatters photons that simultaneously decohere the object and acquire a record of its location. Those photons then move away, guaranteeing the permanence of the record and the consistency of the history of the position of the object.  (In this case, the formally time-dependent Hamiltonian of our simple example is replaced with a time-independent Hamiltonian where the spatial positions of the photons function as clock.) The environment is then effectively renewed, with additional, independent scattering events taking place. The redundancy of records further prevents the past from interfering with the future, as not one, but essentially all the photons would have to return to detect any signature of a superposition -- i.e.,\ Schr\"odinger's cat -- in the historical record. We will return to this example in Ref. \cite{riedel2013objective2}, both elaborating on the points made here and showing how a quantum Darwinism approach enables one to extract the redundant histories.

\section{Conclusion}
\label{sec:thisdiscussion}

Quantum Darwinism accounts for the objective reality of the classical states in our quantum Universe by recognizing that they leave multiple records in the environment. These records can be accessed by observers, providing them with the information about pointer states of the system. As observers do not interact with the system of interest directly, they do not perturb its state while acquiring information.

Decoherence can lead to such dissemination of multiple records throughout the environment. For some environments (e.g., photons) records can be accessed locally in space and time. In other cases (e.g., air) the environment still acquires multiple records, but its subsequent mixing evolution means that they are no longer locally accessible from the natural fragments of the environment  (and, hence, are of no use for observers). In still other cases, the environment may start in a completely mixed state; the information transferred during decoherence is then not accessible locally at all. 

For a pure global state, the consistency of a history is a global property representing orthogonality of the branches.  Partial-trace consistency means the stronger fact that the history is consistent due to orthogonality \emph{in the particular subsystem traced out}. Therefore, partial-trace consistency captures mathematically the intuitive idea that a history has been \emph{decohered} by a fragment of the environment.  When a history is partial-trace consistent with respect to many fragments, it has been \emph{redundantly decohered}.

Quantum Darwinism relies on decohering evolutions that can generate redundant records. In the case of a mixed global state, decoherence does not necessarily imply records.  This realization arose in the contexts of decoherence \cite{Zurek1982}, quantum Darwinism \cite{zwolak2014amplification, Zwolak2009, Zwolak2010, Riedel2010, Riedel2011,Riedel2012,QCBspins}, and consistent histories \cite{Gell-Mann1993, Gell-Mann1998}.  Using the definition of redundancy in the consistent histories framework we have introduced in this paper, we can combine these observations into a single stronger principle: The redundant consistency of histories is a necessary condition for the existence of redundant records, and it is sufficient condition when the global state is pure.

Redundantly recorded histories become ``the objective classical past'' in the same sense that systems take on objective classical states when their pointer basis is redundantly recorded. Many observers can access records of the past independently and simultaneously, and they will arrive at mutually consistent accounts of what occurred. In other words, when they meet and compare their notes, they will be able to agree on the sequence of events that constitutes the history. Although some of the accounts may be incomplete, all accounts will be compatible with the idea that there was just one classical past. 

Quantum Darwinism solves a key interpretational dilemma of quantum theory: It supplies the evidence of objective existence -- here, of the objective past -- using the same mechanism that accounts for (the symptoms of) the objective existence of selected quantum states. This sidesteps metaphysical concerns, such as whether quantum states are epistemic (as some of Bohr's writings may suggest) or ontic (as, one might have thought, would be needed to account for everyday existence), focusing instead on how the evidence of objective classical reality arises in our quantum Universe. Amplification -- which leads to redundancy -- is the key. It singles out collections of sequences of events, encoded as a consistent history, for which consensus is possible. 

The price to be paid for the accessibility of histories is coarse graining; historical records concern selected degrees of freedom of the Universe, but not all degrees of freedom can be recorded. The coarse grainings suggested by quantum Darwinism recognize the role of natural subsystems of the Universe in acquiring information and in decoherence. Moreover, redundant records of these coarse-grained (macroscopic) degrees of freedom are ubiquitous in the classical world we observe.  Redundant consistency is a strong constraint on histories in the sense that, for any given state, the vast majority of consistent sets of histories will not satisfy it. Uniqueness results arising in the study of quantum Darwinism \cite{Ollivier2004,Ollivier2005,horodecki2015quantum, brandao2015generic, zwolak2013complementarity, riedel2013local} can be brought to bear on the set selection problem \cite{Dowker1995, Dowker1996, Kent1996,Kent1997a,Kent1997b, KentRemarks, okon2014measurements, okon2015consistent}, pointing the way toward consistent sets describing the intuitive quasiclassical domain.

The synthesis of consistent histories with quantum Darwinism we started to investigate here shifts the focus from the histories {\it per se} to their evidence broadcast into the world. Thus, rather than study consistency of the sequences of events defining histories stroboscopically -- at discrete instants of time -- via suitable projection operators, we rely on the records of events deposited in the environment. The presence of such records suffices to render a history objective -- simultaneously accessible to many observers. 



\begin{acknowledgments}
We thank Charles Bennett, Robert Griffiths, James Hartle, Adrian Kent, and Lev Vaidman for discussion.  We also thank the University of Ulm for hosting us while this work was being prepared.  This research was partially supported by the U.S. Department of Energy through the LANL/LDRD program, by the John Templeton Foundation, and by the Foundational Questions Institute grant \# 2015-144057 on ``Physics of What Happens''.  Research at the Perimeter Institute is supported by the Government of Canada through Industry Canada and by the Province of Ontario through the Ministry of Research and Innovation.
\end{acknowledgments}

\appendix
\section{Alternate consistent histories}
\label{sec:alt-CNOT-histories}

In this appendix we consider the CNOT example from above and construct alternate sets of consistent histories that fail to be redundantly consistent in order to highlight the strength of redundant consistency as a constraint.

First note how many sets of histories are (merely) consistent but mutually incompatible. \emph{Any} orthonormal basis for the global state, $\ket{\psi} = \sum_{j=1}^J c_j \ket{j}$, (with the states $\{\ket{j}\}$ generically highly entangled with respect to the subsystem tensor product structure) can be used to form the branches of a set of consistent histories.  Indeed, one can define class operators for those histories by grouping the branches together to form an arbitrary tree as in Figure \ref{fig:purestatepartialorder} and then selecting Heisenberg-picture projectors onto the subspaces corresponding to the nodes of the tree. (The corresponding time steps must respect the partial order of the nodes, but are otherwise arbitrary.)

For concreteness, let us look at the CNOT example with two dynamical branching events (instead of three) and with the time steps of the history corresponding to the branching events (i.e., $m=\tilde{m}=2$).   The state at the intermediate ($m=1$) time will be
\begin{align}\begin{split}
\label{FirstBranchState}
\ket{\psi(t_1)} &= U_{\CN}^{\Sys \Env_1} U_{\branch}^\Sys \ket{\psi^0}  \\
&= \frac{1}{\sqrt{2}} \ket{0}_\Sys \ket{\rec=00}_\Env +  \frac{1}{\sqrt{2}} \ket{1}_\Sys \ket{\rec=10}_\Env. 
\end{split}\end{align}
and at the final ($m=2$) time 
\begin{align}\begin{split}
\label{SecondBranchState}
\ket{\psi(t_2)} &= U_{\CN}^{\Sys \Env_2} U_{\branch}^\Sys \ket{\psi(t_1)}  \\
&= \frac{1}{2} \ket{0}_\Sys \ket{\rec=00}_\Env +  \frac{1}{2} \ket{1}_\Sys \ket{\rec=01}_\Env \\
&\,\,\, +  \frac{1}{2} \ket{0}_\Sys \ket{\rec=10}_\Env -  \frac{1}{2} \ket{1}_\Sys \ket{\rec=11}_\Env.
\end{split}\end{align}
We define the fixed, unnormalized vectors
\begin{align}\begin{split}
\label{somevecs}
\ket{A} &= \frac{1}{3\sqrt{2}}\big(2 \ket{0}_\Sys \ket{\alpha=00}_\Env + \ket{1}_\Sys \ket{\alpha=10}_\Env + 2 \ket{0}_\Sys \ket{\alpha=11}_\Env \big) \\
\ket{B} &= \frac{1}{3\sqrt{2}}\big(\ket{0}_\Sys \ket{\alpha=00}_\Env + 2 \ket{1}_\Sys \ket{\alpha=10}_\Env - 2 \ket{0}_\Sys \ket{\alpha=11}_\Env \big) \\
\ket{W} &= \frac{1}{6}\big(2\ket{0}_\Sys \ket{\alpha=00}_\Env + 2 \ket{1}_\Sys \ket{\alpha=01}_\Env + 2 \ket{1}_\Sys \ket{\alpha=10}_\Env \big) \\
\ket{X} &= \frac{1}{6}\big(\ket{0}_\Sys \ket{\alpha=00}_\Env + \ket{1}_\Sys \ket{\alpha=01}_\Env - 2 \ket{1}_\Sys \ket{\alpha=10}_\Env \big) \\
\ket{Y} &= \frac{1}{6}\big(\ket{0}_\Sys \ket{\alpha=10}_\Env - \ket{1}_\Sys \ket{\alpha=11}_\Env + 2 \ket{0}_\Sys \ket{\alpha=11}_\Env \big) \\
\ket{Z} &= \frac{1}{6}\big( 1\ket{0}_\Sys \ket{\alpha=10}_\Env - 2 \ket{1}_\Sys \ket{\alpha=11}_\Env - 2 \ket{0}_\Sys \ket{\alpha=11}_\Env \big)
\end{split}\end{align}
One can check that $\{\ket{A},\ket{B}\}$ and $\{\ket{W},\ket{X},\ket{Y},\ket{Z}\}$ are both orthogonal sets, and that
\begin{align}\begin{split}
\ket{\psi(t_1)} &= \ket{A}+\ket{B}\\
U_{\CN}^{\Sys \Env_2} U_{\branch}^\Sys \ket{A} &= \ket{W} + \ket{Y}\\
U_{\CN}^{\Sys \Env_2} U_{\branch}^\Sys \ket{B} &= \ket{X} + \ket{Z}\\
\ket{\psi(t_2)}&=\ket{W}+\ket{X}+\ket{Y}+\ket{Z}
\end{split}\end{align}
Therefore we can take $\ket{W},\ket{X},\ket{Y},\ket{Z}$ to be our final branches, and construct histories using the Schr\"odinger picture projectors
\begin{align}\begin{split}
P_A^{(1)} &= \projector{A}\\
P_B^{(1)} &= \projector{B} \\
P_{\overline{AB}}^{(1)} &= I - P_A^{(1)} - P_B^{(1)}.
\end{split}\end{align}
and
\begin{align}\begin{split}
P_W^{(1)} &= \projector{W} \\
P_X^{(1)} &= \projector{X} \\
P_Y^{(1)} &=\projector{Y} \\
P_Z^{(1)} &=\projector{Z} \\
P_{\overline{WXYZ}}^{(1)} &= I - P_W^{(1)} - P_X^{(1)} - P_Y^{(1)} - P_Z^{(1)}.
\end{split}\end{align}
The class operator of history $\alpha = BZ$ would be $C_{BZ} = P_B^{(1)} (t_1) P_Z^{(2)} (t_2)$, etc.  There are four non-zero histories with associated probabilities, corresponding to the four branches ($\ket{W},\ket{X},\ket{Y},\ket{Z}$), but they do not have a useful physical interpretation. Likewise, the projectors $\projector{W},\projector{X},\projector{Y},$ and $\projector{Z}$ are formal records of the corresponding histories (in the sense of Gell-Mann and Hartle), but they are not useful to observers who are unable to make arbitrary non-local measurements.\footnote{Note that these (unphysical) records differ from the corresponding class operators, e.g., $\projector{Z} \neq C_{BZ}$.}

We can also construct a set of histories that is $\Env$-consistent (and therefore consistent) but not redundantly consistent.  First, define the unnormalized states
\begin{align}\begin{split}
\ket{\theta}_\Env &= (\cos \theta +  \sin \theta) (\cos \theta \ket{\rec=00}_\Env + \sin \theta \ket{\rec=10}_\Env) \\ 
\ket{\bar{\theta}}_\Env &= (\sin \theta - \cos \theta) (\sin \theta \ket{\rec=00}_\Env - \cos \theta \ket{\rec=10}_\Env) \label{rotst}
\end{split}\end{align}
for some angle $\theta$.  Note that $\braket{\theta}{\bar{\theta}}_\Env = 0$ and $\ket{\theta}_\Env + \ket{\bar{\theta}}_\Env  = \ket{\rec=00}_\Env + \ket{\rec=10}_\Env$.  Then, with a similar definition (differing only in a minus sign) using an arbitrary angle $\phi$ and the states $\ket{\rec=01}_\Env$ and  $\ket{\rec=11}_\Env$, we can take our branches to be the orthogonal set
\begin{align}\begin{array}{llll}
\ket{0}_\Sys \ket{\theta}_\Env, \ket{0}_\Sys \ket{\bar{\theta}}_\Env, \ket{1}_\Sys \ket{\phi}_\Env, \ket{1}_\Sys \ket{\bar{\phi}}_\Env.
\end{array}\end{align}
Together these sum up to the global state in Eq.~\eqref{SecondBranchState}.  Just as in the previous example, we can choose an arbitrary set of ordered time steps and define projectors (making sure to complete the basis at intermediate time steps) and class operators of histories for which the above states are branches.  

Again, to be more concrete, consider the projectors (appropriately normalized and at each time step ignoring the members of the complete set that play no role) at the first 
\begin{align}\begin{split}
P_1^{(1)} &= \projector{0,\theta} + \projector{0,\bar{\theta}} \\
P_2^{(1)} &=\projector{1, \theta} + \projector{1,\bar{\theta}} \\
\end{split}\end{align}
and second
\begin{align}\begin{split}
P_1^{(2)} &= \projector{0,\theta} \\
P_2^{(2)} &= \projector{0,\bar{\theta}} \\
P_3^{(2)} &= \projector{1,\phi} \\
P_4^{(2)} &= \projector{1,\bar{\phi}} 
\end{split}\end{align}
time steps. This set of histories is not only consistent, it is $\Env$-consistent, since all the branches lie in orthogonal subspaces of the environment.  This is true regardless of the choice of angles. Thus, we can construct an infinite number of sets of branches that give consistent or partial-trace consistent histories. The freedom exploited in Eq. (\ref{rotst}), however, is not present for redundantly consistent histories, hence excluding them and narrowing down the set selection problem. As indicated in the main text, there is only one set of branches that are redundantly consistent.

\bibliographystyle{apsrev4-1}
\bibliography{riedelbib}
\end{document}